\documentclass[sn-mathphys,Numbered]{sn-jnl}% Math and Physical Sciences Reference Style
\usepackage{graphicx}%
\usepackage{multirow}%
\usepackage{amsmath,amssymb,amsfonts}%
\usepackage{amsthm}%
\usepackage{mathrsfs}%
\usepackage[title]{appendix}%
\usepackage{xcolor}%
\usepackage{textcomp}%
\usepackage{manyfoot}%
\usepackage{booktabs}%
\usepackage{algorithm}%
\usepackage{algorithmicx}%
\usepackage{algpseudocode}%
\usepackage{listings}%

\usepackage{array}
\usepackage[caption=false,font=normalsize,labelfont=sf,textfont=sf]{subfig}
\usepackage{textcomp}
\usepackage{stfloats}
\usepackage{verbatim}
\theoremstyle{thmstyleone}%
\theoremstyle{thmstyletwo}%

\theoremstyle{thmstylethree}%

\begin{document}

\title[The impact of spatio-temporal travel distance on epidemics using an interpretable attention-based sequence-to-sequence model]{The impact of spatio-temporal travel distance on epidemics using an interpretable attention-based sequence-to-sequence model}


\author[1]{\fnm{Yukang} \sur{Jiang}}
\equalcont{These authors contributed equally to this work.}

\author[1]{\fnm{Ting} \sur{Tian}}
\equalcont{These authors contributed equally to this work.}

\author[1]{\fnm{Huajun} \sur{Xie}}
\equalcont{These authors contributed equally to this work.}

\author*[2]{\fnm{Hailiang} \sur{Guo}}\email{2316384027@bjmu.edu.cn}
\author*[3]{\fnm{Xueqin} \sur{Wang} }\email{wangxq20@ustc.edu.cn}
%\author*[2]{\fnm{Hailiang} \sur{Guo}}\email{ }
%\author*[3]{\fnm{Xueqin} \sur{Wang} }\email{ }
%\thanks{Corresponding authors: Xueqin Wang; Hailiang Guo}

\affil[1]{\orgdiv{School of Mathematics}, \orgname{Sun Yat-sen University}, \orgaddress{\city{Guangzhou}, \postcode{510275}, \country{China}}}

\affil*[2]{\orgdiv{School of Public Health}, \orgname{Peking University}, \orgaddress{\city{Beijing}, \postcode{100871}, \country{China}}}

\affil*[3]{\orgdiv{School of Management}, \orgname{University of Science and Technology of China}, \orgaddress{\city{Hefei}, \postcode{230026}, \country{China}}}


\abstract{Amidst the COVID-19 pandemic, travel restrictions have emerged as crucial interventions for mitigating the spread of the virus. In this study, we enhance the predictive capabilities of our model, Sequence-to-Sequence Epidemic Attention Network (S2SEA-Net), by incorporating an attention module, allowing us to assess the impact of distinct classes of travel distances on epidemic dynamics. Furthermore, our model provides forecasts for new confirmed cases and deaths. To achieve this, we leverage daily data on population movement across various travel distance categories, coupled with county-level epidemic data in the United States. Our findings illuminate a compelling relationship between the volume of travelers at different distance ranges and the trajectories of COVID-19. Notably, a discernible spatial pattern emerges with respect to these travel distance categories on a national scale. We unveil the geographical variations in the influence of population movement at different travel distances on the dynamics of epidemic spread. This will contribute to the formulation of strategies for future epidemic prevention and public health policies.}




\maketitle

\section{Introduction}\label{sec1}
Since the commencement of the 21st century, the salience of infectious diseases within the ambit of public health discourse has burgeoned \cite{frankish2003death, daszak2000emerging, donnelly2003epidemiological}, and we have borne witness to several profound pandemics that have exerted significant societal and economic impacts \cite{hall2020pandemics}, with COVID-19 serving as a prominent illustration. Although the peak of the pandemic has passed, its long-lasting impact remains significant. Reflecting on COVID-19 and its consequences is instrumental in future public health preparedness and control efforts. 

Since March 2020, the world has been grappling with the global pandemic of the novel coronavirus disease (COVID-19), declared by the World Health Organization (WHO) \cite{WHO}. This unprecedented crisis has continued to wreak havoc across the globe, with far-reaching consequences. By November 13, 2021, the United States alone had reported over 47 million cumulative confirmed cases and 0.76 million total deaths, accounting for a staggering one-fifth of all global confirmed cases and one-seventh of total global deaths \cite{NYT}. In response to this dire situation, public health authorities around the world initiated stringent measures in the early stages of the pandemic to curb its spread. For instance, on March 19, 2020, California implemented a ``stay home except for essential needs" order \cite{tian_tan_jiang_wang_zhang_2021}. However, as time passed, we witnessed a gradual resurgence in mobility, approaching levels reminiscent before SARS-CoV-2 emerged \cite{chen2022longitudinal}.

According to data from the Bureau of Transportation Statistics (available at: \url{https://www.bts.gov/daily-travel}), people engage in various travel distances when not confined to their homes, spanning from trips shorter than 1 mile to journeys exceeding 500 miles. In the years 2019 and 2020, more than 16 billion and approximately 12 billion individuals embarked on journeys, respectively. Nevertheless, as of December 2021, the total number of trips had stabilized at around 15.2 billion. To facilitate a clearer understanding of these travel patterns, we have grouped the diverse travel distances into four distinct classes, based on three geographical scales, 1 mile, 50 miles, and 250 miles. These categories encompass trips shorter than 1 mile, those spanning between 1 and 50 miles, journeys between 50 and 250 miles, and finally, trips exceeding 250 miles. We refer to these as ``community-level", ``county-level", ``in-state-level", and ``out-state-level" travel distances, respectively \cite{chen2022longitudinal}. In this context, our study meticulously observes the daily fluctuations in these travel patterns across the United States, spanning the period from 2019 to 2021.

\begin{figure*}[!htb]
\centering
\includegraphics[scale = 0.5]{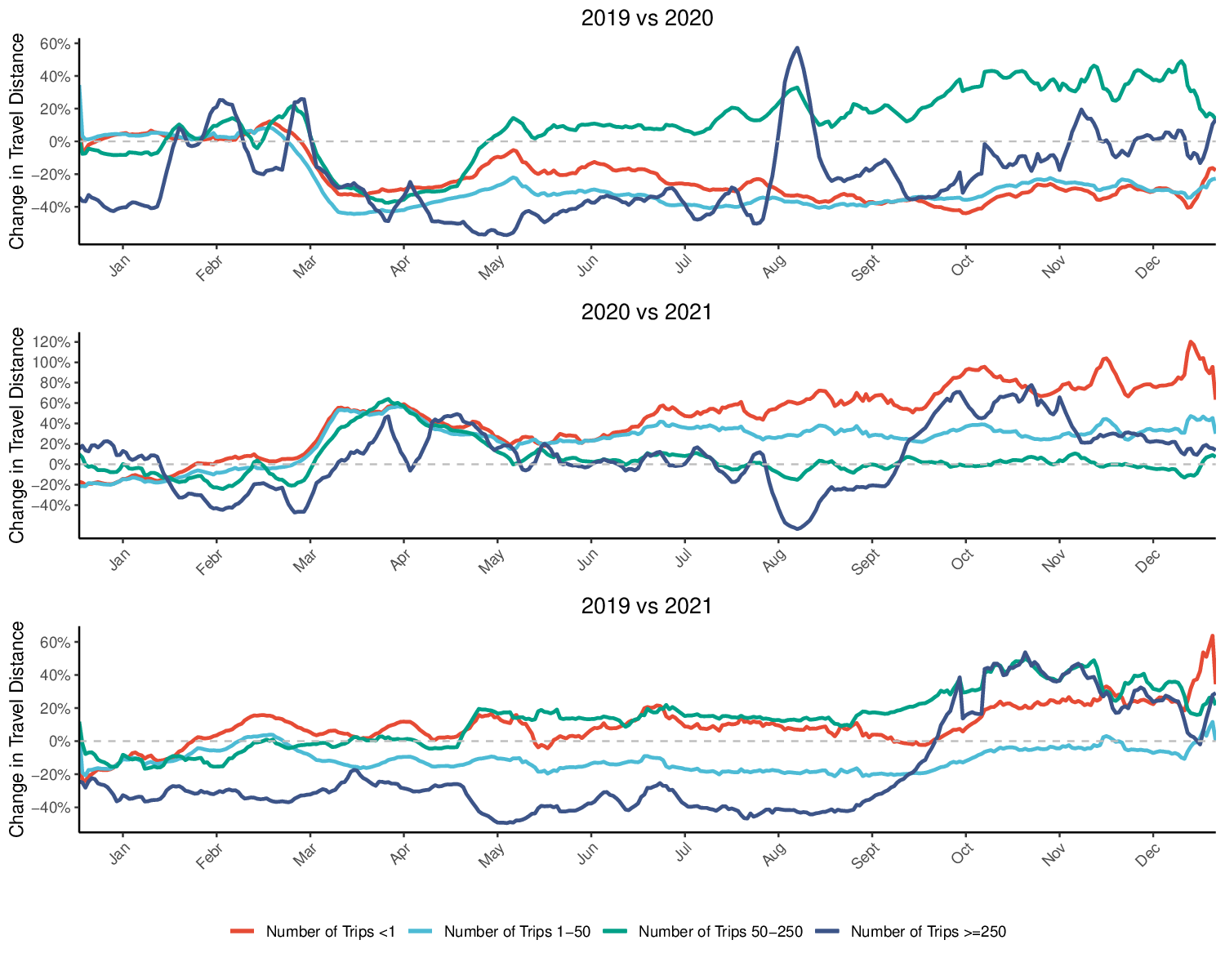}
\caption{Daily relative changes in nationwide travel distances (2019-2021) in the United States.} \label{distance}
\end{figure*}

Figure \ref{distance} illustrates the daily fluctuation in travel patterns for various travel distances from 2019 to 2021. The change in travel distance is quantified as the relative difference in the daily number of individuals traveling within each travel distance class, compared to the corresponding periods in 2019, 2020, and 2021. Notably, starting from March 2020, the number of people taking trips across all travel distances began to decline. However, a distinct trend emerged for ``county-level" travel distances (ranging from 50 to 250 miles) as they gradually increased after May 2020 when compared to the same periods in 2019. Intriguingly, the number of people taking ``community-level" trips (distances less than 1 mile) experienced a nearly year-long decline, from March 2020 to February 2021, before returning to pre-pandemic levels after February 2021. This suggests that people progressively resumed their daily routines during 2021, reflecting a unique nationwide change pattern in travel behavior across various distances.

This change in travel distance serves as a proxy for mobility shifts and the effectiveness of travel restrictions in slowing the transmission of the virus. Numerous studies have emphasized that reducing mobility can effectively curb transmission by lowering the effective reproduction number of the virus \cite{flaxman2020estimating,dehning2020inferring,jia2020population}. For instance, Chinazzi et al. \cite{chinazzi2020effect} simulated different levels of travel reduction to assess their impact on mitigating the epidemic in China. Additionally, researchers have constructed mobility networks to model lockdown-induced changes in Germany \cite{schlosser2020covid} and simulate the effects of movement restrictions in France, Italy, and the UK \cite{galeazzi2021human}. Both epidemiological models and mobility networks underscore the benefits of reducing mobility. However, a critical question remains unanswered: which class of travel distance has the most significant impact on epidemic spread, and does the spatio-temporal distribution of travel distance affect both confirmed cases and deaths? The daily travel data, which records various travel distances at the county level throughout the United States, provides a unique opportunity to model the influence of different travel distances on epidemic trajectories.

In the realm of emerging science, deep learning has demonstrated remarkable performance across various domains \cite{ronneberger2015u, ren2015faster, hochreiter1997long, sutskever2014sequence}. In the context of COVID-19 prediction, several sequence-based deep learning methods have achieved high prediction accuracy \cite{pang2021collaborative, zhang2021seq2seq, Tian2020}. For instance, Alassafi et al. \cite{alassafi2022time} compared and evaluated the performance of different prediction models for forecasting COVID-19 cases, recoveries, and deaths, achieving high accuracy in 7-day predictions. However, most of these methods rely solely on sequential deep learning models to predict future epidemic trajectories, which lack interpretability \cite{chandra2022deep, sinha2022analysis}. Moreover, these approaches often overlook the geographic information, which significantly influences prediction and analysis results \cite{giuliani2020modelling, Tian2020}. Furthermore, existing deep learning studies typically analyze and forecast at the state or national level, without delving into smaller county-level modeling \cite{chandra2022deep, sinha2022analysis}. 

In light of these limitations, our study introduces county-level travel distance patterns into sequence-based deep learning to elucidate the qualitative effects of population mobility on epidemic spread. We propose an encoder-decoder model to elucidate the interplay between different travel distances and the spatio-temporal spread of COVID-19. To offer a data-driven approach that not only enhances sequence forecasting accuracy but also provides interpretable insights into travel patterns, we employ the sequence-to-sequence framework \cite{sutskever2014sequence} coupled with attention mechanisms \cite{fukui2019attention}. It takes into account the sequence of epidemic trajectories, historically confirmed cases and deaths, and the number of people traveling at various distances as inputs. Additionally, we incorporate geographical information, including latitude and longitude \cite{Tian2020}, and state-level information to predict the future trajectory of COVID-19 cases and deaths. Our primary objective is to uncover the relationship between travel patterns and new cases and deaths, with a specific focus on understanding which class of travel distance plays a more significant role in driving epidemic spread. Consequently, our work aims to provide an intuitive understanding of travel patterns and their impact on predicting and interpreting the course of the epidemic across different time periods and regions of the United States. This information holds significant implications for policymakers striving to refine and tailor COVID-19 mobility regulations.

\section{Method}\label{sec2}
\subsection{Data Sources}
We collected daily records of new confirmed COVID-19 cases and deaths, commencing from the day when patients were first confirmed in each affected county within the United States, and continued this data compilation until November 13, 2021. The primary source for our epidemic data is The New York Times \cite{NYT}. Concurrently, we gathered daily data pertaining to the population's travel behavior, categorizing it into four distinct classes based on travel distances. This travel data was collected contemporaneously with the epidemic data and can be accessed at the following link: \url{https://www.bts.gov/daily-travel} from all 3,118 counties in the 50 states of the United States. Additionally, we extracted geographical information, including longitude and latitude, for each affected county from the Census TIGER 2000 dataset \cite{Counties}.

\subsection{Data Utilization}
In our endeavor to evaluate the efficacy of our proposed model, we adopted a randomized division of the 3,118 counties into two sets: training counties and test counties, maintaining a ratio of 4:1 between them. We initiated the validation of our model from the very first day when confirmed cases were officially reported within a county in the United States, continuing this validation process until November 13, 2021. Consequently, we projected the trajectories of COVID-19 cases and deaths for the last 7 days and 30 days leading up to November 13, 2021.

\begin{figure*}[!htb]
\centering
\includegraphics[scale = 0.5]{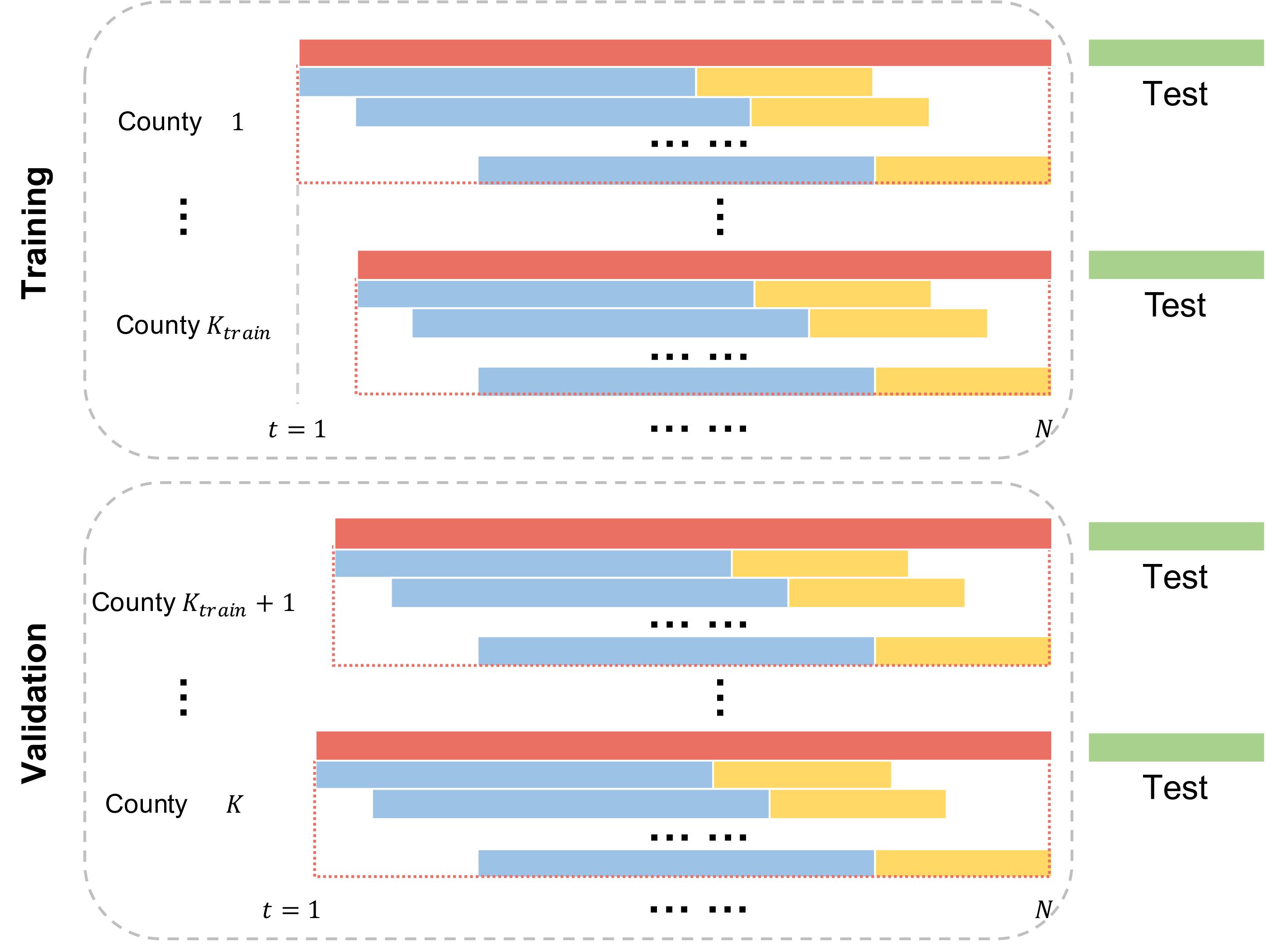}
\caption{The data utilization structure in our proposed model.} \label{structure}
\end{figure*}

To facilitate the understanding of our data utilization process, we present a graphical representation in Figure \ref{structure}. In this representation, each county's data is illustrated by a combination of red and green bars. These bars collectively depict the entire duration from the day when cases were first reported in a county in the United States to November 13, 2021. Blue bars within this representation correspond to observed data, encompassing cases, deaths, and data related to the four classes of travel distances, captured over 30-day intervals. The yellow bars represent the subsequent 7-day or 30-day projections for cases and deaths within a county. The final yellow bar signifies the conclusion of the last 7-day or 30-day data point for COVID-19 before November 13, 2021. While the commencement times vary for each county's data, the termination point is the same, resulting in different total input data lengths for each county.

The expression for the inputs of the epidemic data ($E$) is expressed as:
\begin{equation*}
\textbf{X}_{..k}^{(E)}=\left( \begin{matrix} 
x_{1,k}^{(cases)}&\cdots &x_{30,k}^{(cases)} & x_{1,k}^{(deaths)} &\cdots &x_{30,k}^{(deaths)} \\
\vdots&\cdots&\cdots&\cdots&\cdots&\vdots \\
x_{N-37,k}^{(cases)}&\cdots &x_{N-8,k}^{(cases)} &x_{N-37,k}^{(deaths)} &\cdots &x_{N-8,k}^{(deaths)}
\end{matrix}
\right)_{T\times60}, \quad k=1,2,...K,
\end{equation*}
where $N$ is the length of the training period, $T = N-30$, and $K$ is the total number of counties. $x_{i,k}^{(cases)}$ represents the new confirmed cases, and $x_{i,k}^{(deaths)}$ represents the new deaths on the corresponding date. For instance, when $i=1$, it corresponds to the first day when the confirmed cases were officially reported in a county. These new confirmed cases and deaths give rise to 60 historical epidemic variables as the first part of the input data, which form the first part of the input data.

Besides, we define the travel distance data ($D$) as:
\begin{align*}
\textbf{D}_{..k}&=\left( \begin{matrix} 
d_{1,1,k} &d_{1,2,k} & d_{1,3,k} &d_{1,4,k} \\
\vdots&\vdots&\vdots&\vdots \\
d_{N,1,k}&d_{N,2,k} &d_{N,3,k}  &d_{N,4,k}
\end{matrix}
\right)_{N\times4}, \\ k&=1,2,...K.
\end{align*}
where the columns of $\textbf{D}_{..k}$ represent the number of individuals taking trips at different travel distance categories: ``community-level" (less than 1 mile), ``county-level" (between 1 mile and 50 miles), ``in-state-level" (between 50 miles and 250 miles), and ``out-state-level" (over 250 miles). These four travel distance variables constitute the second part of the input data.

For a 7-day prediction, the corresponding predicted cases and deaths are represented as:
\begin{align*}
\textbf{Y}_{k}&=\left( \begin{matrix} 
x_{31,k}^{(cases)}&\cdots &x_{37,k}^{(cases)} & x_{31,k}^{(deaths)} &\cdots &x_{37,k}^{(deaths)} \\
\vdots&\cdots&\cdots&\cdots&\cdots&\vdots \\
x_{N-7,k}^{(cases)}&\cdots &x_{N,k}^{(cases)} &x_{N-7,k}^{(deaths)} &\cdots &x_{N,k}^{(deaths)}
\end{matrix}
\right)_{T\times14}, \quad k=1,2,...K.
\end{align*}

\subsection{S2SEA-Net}
In our approach, we utilize the past 30-day epidemic data, which includes new cases and deaths, in conjunction with the corresponding 30-day travel distance data to predict future trajectories of COVID-19 for both 7 and 30-day intervals. Our fundamental model employs a sequence-to-sequence framework based on an encoder-decoder structure, with Long Short-Term Memory (LSTM) \cite{hochreiter1997long} as the base building block. The LSTM structure consists of a cell state $c_t$, an input gate $in_t$, a forget gate $f_t$, and an output gate $o_t$. Given a sequence of multivariate variables $\{x_1,x_2,...,x_t\}$, where $x_t\in\mathbb{R}^d$, we calculate the hidden state $h_t$ as follows:
$$
\begin{aligned}
&in_t = \sigma(W_i x_t+U_i h_{t-1}+b_i)\\
&f_t = \sigma(W_f x_t+U_f h_{t-1}+b_f)\\
&o_t = \sigma(W_o x_t+U_o h_{t-1}+b_o)\\
&\tilde{c}_t = \tanh(W_c x_t+U_c h_{t-1}+b_c)\\
&c_t = f_t\odot c_{t-1} + i_t \odot \tilde{c}_t\\
&h_t = o_t\odot \tanh(c_t)
\end{aligned}
$$
where $\sigma$ and $\odot$ denote the sigmoid function and element-wise multiplication, respectively. In this context, $x_t$ is defined as $\operatorname{Concat}(\textbf{X}_{t.k}^{(E)}, \operatorname{Attention}(D_{t.k}))$, with $\operatorname{Concat}$ representing the concatenation process in deep learning and $\operatorname{Attention}$ being the attention module, which will be discussed shortly.

We then construct a sequence-to-sequence framework, which falls under the broader category of the encoder-decoder structure. This framework can transform one sequence into another sequence \cite{sutskever2014sequence}, allowing for different lengths of input and output sequences. While previous studies have employed sequence-to-sequence architectures for COVID-19 forecasting \cite{pang2021collaborative, zhang2021seq2seq}, their focus primarily centered on predicting epidemic trends and lacked specific information for decision-makers. In our approach, we incorporate a designed attention module into our basic model, denominated as the Sequence-to-Sequence Epidemic Attention Network (S2SEA-Net), to quantify the impact of travel distances on the spread of COVID-19.

The attention module serves to learn the weights assigned to each variable and each time step, denoted as $\alpha_t \in \mathbb{R}^{30\times4}$. This module combines travel distance data with geographical information, including latitude, longitude, and state information (state ID). The categorical variable representing state ID is transformed using an embedding layer for feature representation, while the continuous variables for latitudes and longitudes are processed through a linear layer. Each geographical variable is represented by an $h$-dimensional vector. These representations are then combined using concatenation operation to form a global geographical information representation, denoted as $g_k$, where $k = 1, \cdots, K$, and each county may have a unique representation.

Each sample of travel distance data, $\mathbf{d}_t \in \mathbb{R}^{30\times4}$, consists of 30 rows representing 30-day periods and 4 columns representing the number of people traveling for four classes of travel distances. This data is transformed using a linear layer to create representations and is then flattened to
$$
\mathbf{d}_t^{(r)}=\left( \begin{matrix} 
d_{t,1}^{(r)\top} \\
\vdots \\
d_{t,30}^{(r)\top}
\end{matrix}
\right)_{30\times (4h)}.
$$

We combine the geographical information by
\begin{align*}
\eta_{t, j} &= \operatorname{BN}{(W_{d} \cdot d_{t, j}^{(r)}+b_{d})} + \operatorname{BN}{(W_g \cdot (g_k) + b_g)}, \\
\quad j &= 1, \cdots, 30, \quad t = 1, \cdots, N - 30,
\end{align*}
where $W_{d}, W_g \in \mathbb{R}^{4\times4h}$ are the weights, and $b_{d}, b_g \in \mathbb{R}^{4 \times 1}$ are constants. $\operatorname{BN}$ refers to a batch normalization layer.

Subsequently, the weights of the population taking trips for each class of travel distance are normalized and summed to $1$. We apply the ``elu" and ``tanh" activation functions to enhance the model's expression capacity after the linear layer. The ``softmax" activation function is then used to constrain the outputs between $0$ and $1$. The attention weights are computed as:
$$
\alpha_{t, j} = \operatorname{softmax}(f(\eta_{t, j})),\quad j=1,2,\dots,30,
$$
where $f$ consists of a linear layer and $\tanh$ activation function. The formula below is used to aggregate the travel distance data and calculate the context vector $v=[v_1,v_2,\dots,v_{30}]$:
$$
v_{t,j} = \sum^{4}_{i}\alpha_{t,j,i}d_{t,j,i}.
$$

\begin{figure*}[!htb]
\centering
\includegraphics[scale = 0.25]{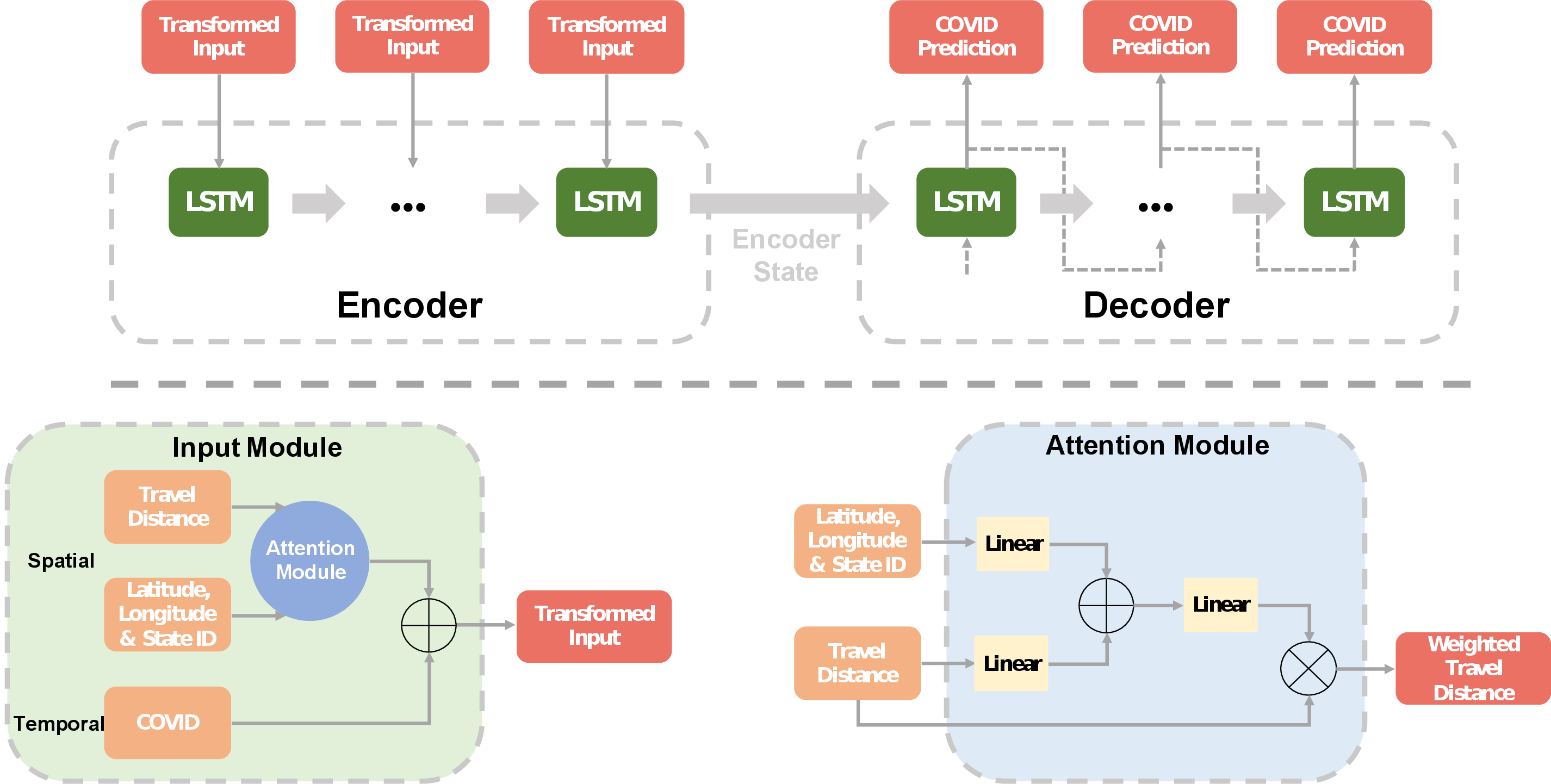}
\caption{The structure of the S2SEA-Net. The upper panel represents the primary model with an encoding and decoding structure, while the lower panels depict the input and attention modules, respectively.} \label{method}
\end{figure*}

After $v_{t,j}$ are obtained, we concatenate it with $\textbf{X}^{(E)}_{..k}$ as the transformed data that enters the encoder. Subsequently, we predict $\textbf{Y}_k$ in the decoder. Consequently, every 30-day epidemic and travel distance dataset becomes the transformed input in the encoder, and the decoder generates the 7- and 30-day predicted epidemic trajectories. The overall structure of the proposed deep neural network is illustrated in Figure \ref{method}, and we refer to this model as ``S2SEA-Net".

\subsection{Implementation details}
We implemented the S2SEA-Net using Keras 2.6. The following details provide a comprehensive overview of the network configurations and training procedures for each module within the S2SEA-Net:
\begin{itemize}
\item In the LSTM layer, there are 256 nodes, and dropout regularization with a probability of 0.2 is applied. The node count and dropout regularization details are consistent with Tian et al. \cite{Tian2020}.
\item The dimension ($h$) of the hidden vector in the attention module is set to 64.
\item During training, we use stochastic gradient descent (SGD) with an initial learning rate of 0.1 and a stepped decay rate of 0.1 every 40 epochs, for a total of 200 epochs. To prevent overfitting, we employ the early stopping strategy.
\item The loss function employed is root-mean-square error (RMSE), used to assess the accuracy of the S2SEA-Net:
\[RMSE = \sqrt{\frac{\sum^t_{i=1}(Predicted_i-Observed_i)^2}{t}}, \text{ where } t=7,30.\]
Here, $Predicted_i$ represents the predicted new confirmed cases or new deaths on the $i^{th}$ day, and $Observed_i$ represents the observed values on the corresponding date.
\end{itemize}

\section{Results}

\subsection{Spatial effects of travel distances on COVID-19}

\begin{table*}[!htb]
\caption{The different 10 counties with the relatively larger average weights over the entire period for the population taking trips for each of four classes of travel distances.}
\resizebox{\textwidth}{!}{%
\begin{tabular}{clc}
\hline
\textbf{Travel distances}                                                       & \textbf{Counties}                                & \textbf{Weights} \\ \hline
\multirow{10}{*}{\textbf{``Community-level"   Trips ($<$ 1 mile)}}              & California Riverside                             & 0.7002           \\
                                                                                & California San Bernardino                        & 0.6943           \\
                                                                                & \textbf{California Los Angeles} & 0.6658           \\
                                                                                & Texas Harris                                     & 0.6198           \\
                                                                                & California San Diego                             & 0.5970            \\
                                                                                & California San Joaquin                           & 0.5944           \\
                                                                                & Texas Bexar                                      & 0.5937           \\
                                                                                & California Orange                                & 0.5807           \\
                                                                                & California Kern                                  & 0.5792           \\
                                                                                & Texas Tarrant                                    & 0.5536           \\ \hline
\multirow{10}{*}{\textbf{``County-level"   Trips (between 1 and 50 miles)}}     & \textbf{Hawaii Honolulu}        & 0.3489           \\
                                                                                & Alaska Nome Census Area                          & 0.3351           \\
                                                                                & Alaska Kodiak Island Borough                     & 0.3286           \\
                                                                                & Washington Wahkiakum                             & 0.3243           \\
                                                                                & Alaska Bethel Census Area                        & 0.3242           \\
                                                                                & Alaska Fairbanks North Star Borough              & 0.3211           \\
                                                                                & Washington Jefferson                             & 0.3205           \\
                                                                                & Washington Skamania                              & 0.3203           \\
                                                                                & Alaska Yukon-Koyukuk Census Area                 & 0.3203           \\
                                                                                & Washington Pacific                               & 0.3196           \\ \hline
\multirow{10}{*}{\textbf{``In-state-level"   Trips (between 50 and 250 miles)}} & \textbf{Illinois Cook}          & 0.5354           \\
                                                                                & Alaska Fairbanks North Star Borough              & 0.5148           \\
                                                                                & Alaska Southeast Fairbanks Census Area           & 0.5146           \\
                                                                                & Alaska Yukon-Koyukuk Census Area                 & 0.5146           \\
                                                                                & Alaska Matanuska-Susitna Borough                 & 0.5145           \\
                                                                                & Alaska Anchorage                                 & 0.5084           \\
                                                                                & Alaska Juneau City and Borough                   & 0.5058           \\
                                                                                & Alaska Nome Census Area                          & 0.4989           \\
                                                                                & Alaska Kenai Peninsula Borough                   & 0.4986           \\
                                                                                & Alaska Petersburg Borough                        & 0.4967           \\ \hline
\multirow{10}{*}{\textbf{``Out-state-level"   Trips ($>$ 250 miles)}}           & Florida Glades                                   & 0.3740            \\
                                                                                & Florida Monroe                                   & 0.3734           \\
                                                                                & Florida Hendry                                   & 0.3713           \\
                                                                                & Florida Okeechobee                               & 0.3701           \\
                                                                                & Florida Hardee                                   & 0.3700             \\
                                                                                & Florida DeSoto                                   & 0.3698           \\
                                                                                & Florida Martin                                   & 0.3627           \\
                                                                                & Florida Highlands                                & 0.3627           \\
                                                                                & \textbf{Florida Indian River}   & 0.3611           \\
                                                                                & Florida Charlotte                                & 0.3608           \\ \hline
\end{tabular}}
\label{table2}
\end{table*}

Table \ref{table2} provides an investigation into how average weights in the number of the population taking trips for four classes of travel distances vary with geography. Notably, 7 counties in California have fairly higher average weights for ``community-level" trips (very short travel distances), while 10 counties in Florida have relatively higher average weights for ``out-state-level" trips (long travel distances). Additionally, 9 counties in Alaska have moderately higher average weights for ``in-state-level" trips (intermediate travel distances). Overall, the top 10 average weights for the population taking ``community-level" trips are considerably greater than the remaining classes of travel distances.

\begin{figure*}[!htb]
\centering
	\subfloat[California, Los Angeles County]{\includegraphics[width = 0.45\textwidth]{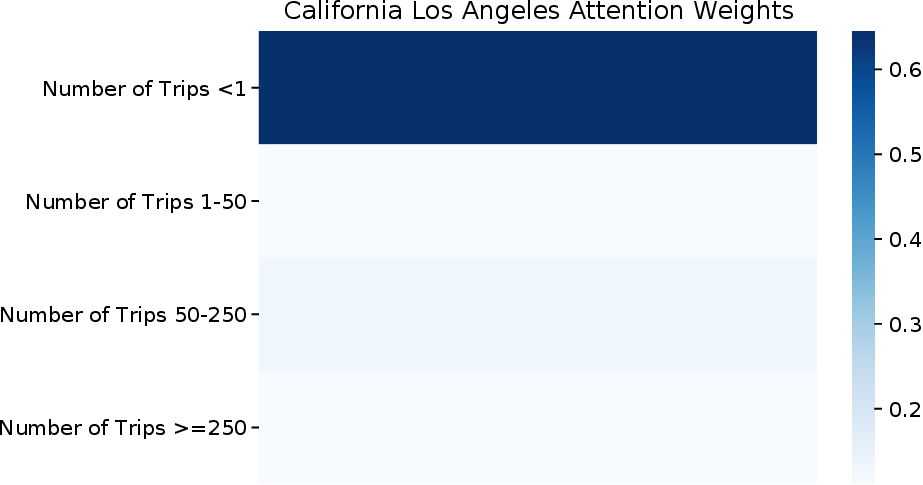}}
	\hfill
	\subfloat[Hawaii, Honolulu County]{\includegraphics[width = 0.45\textwidth]{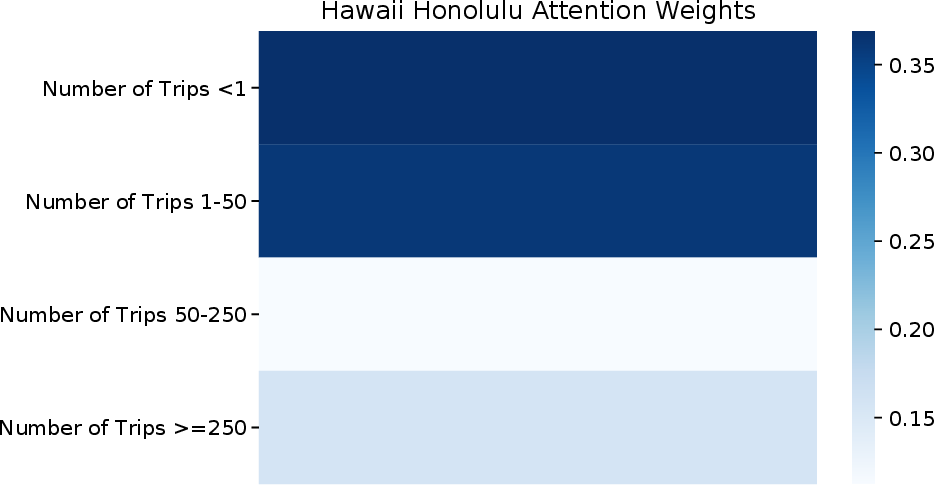}}
	\hfill
	\subfloat[Illinois, Cook County]{\includegraphics[width = 0.45\textwidth]{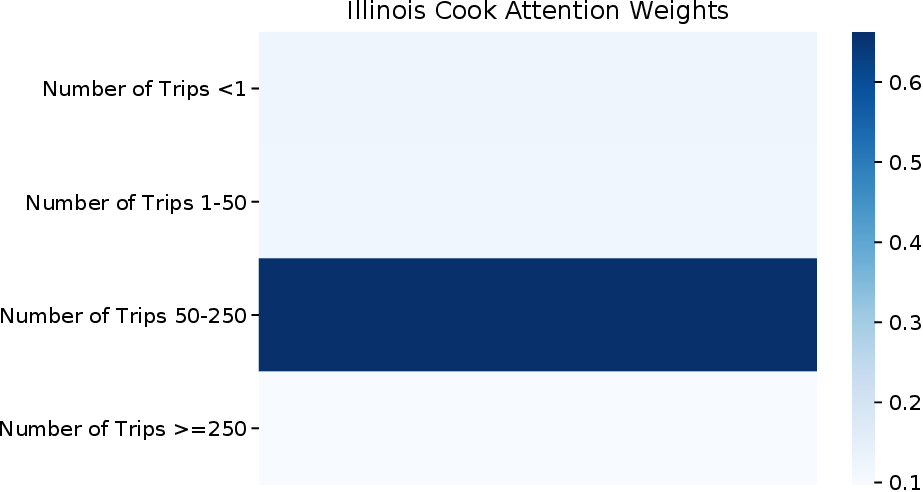}} 
	\hfill
	\subfloat[Florida, Indian River County]{\includegraphics[width = 0.45\textwidth]{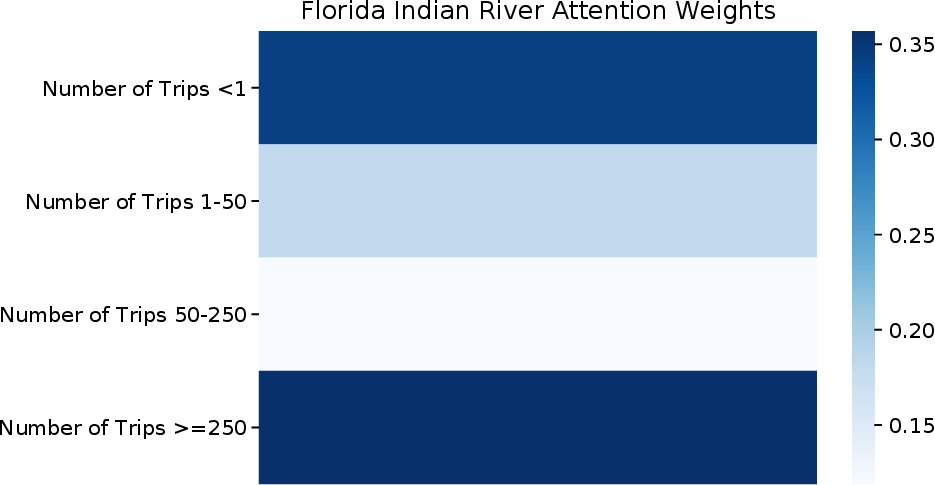}} 
\caption{The heatmap of the average weights over the last 30-day before November 13, 2021 for a full range of travel distances in California Los Angeles, Hawaii Honolulu, Illinois Cook, and Florida Indian River.}
\label{weights}
\end{figure*}

Figure \ref{weights} examines the average weights of four representative counties over the last 30 days before November 13, 2021, for a full range of the population taking trips. These counties exhibit varying patterns in the average weights of different classes of travel distances. For example, California, Los Angeles County, has significantly higher average weights for ``community-level" trips compared to other classes. In contrast, Hawaii, Honolulu County, shows roughly equal average weights for two classes of travel distances. These variations highlight differences in the spatial distribution of travel distances across infected counties.

\begin{figure*}[!htb]
\begin{center}
\includegraphics[scale = 0.5]{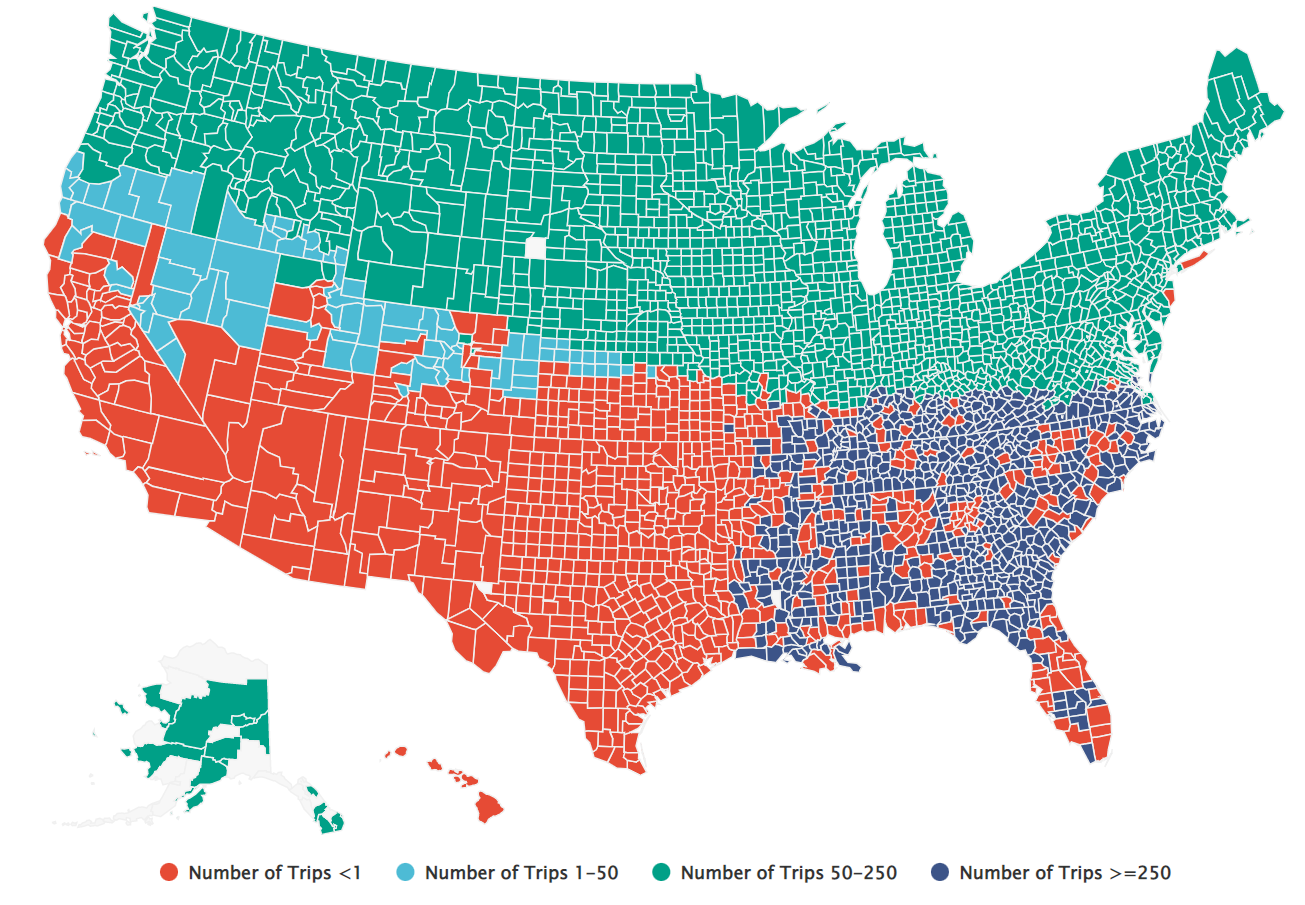}
\end{center}
\caption{The map of the spatial distribution of average weights over the entire period for a full range of travel distances in 3118 infected counties of the United States.} \label{map}
\end{figure*}

For the overall spatial distribution of travel distances in the 3118 infected counties of the United States (Figure \ref{map}), the estimated average weights over the entire period are conceivably varied with geography. For example, the infected counties in the northern region have relatively larger average weights for the population taking ``in-state-level" trips, except for Ocean County in New Jersey and Suffolk County in New York. These two counties and most south-western region counties have relatively larger average weights for the population taking ``community-level" trips. For each class of travel distance, the average weights over the entire period for each infected county have class-specific changes and are spatially different.   

\subsection{Temporal effects of travel distances on COVID-19}

\begin{figure*}[!htb]
\centering
	\subfloat[California, Los Angeles County]{\includegraphics[width = 0.45\textwidth]{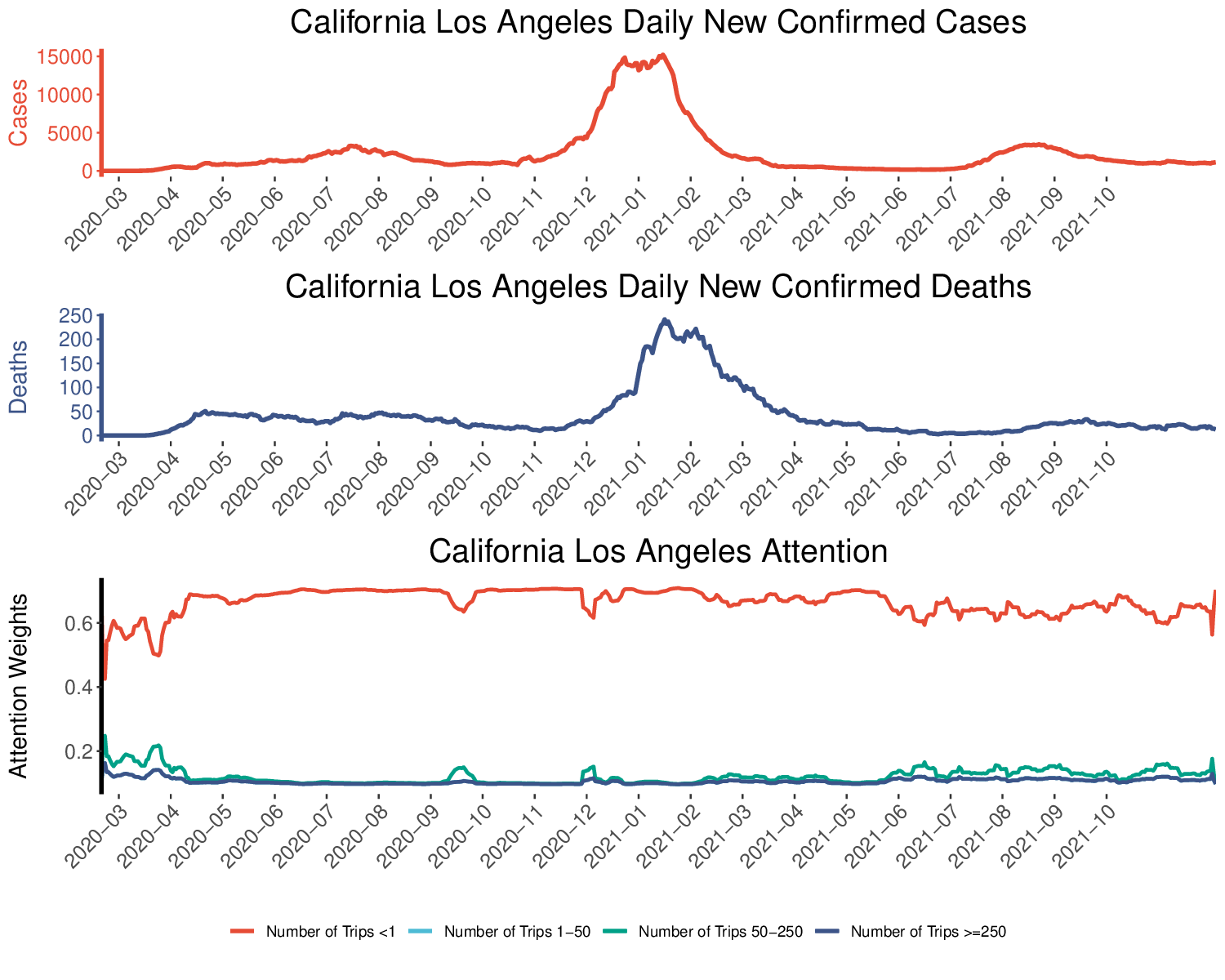}}
	\hfill
	\subfloat[Hawaii, Honolulu County]{\includegraphics[width = 0.45\textwidth]{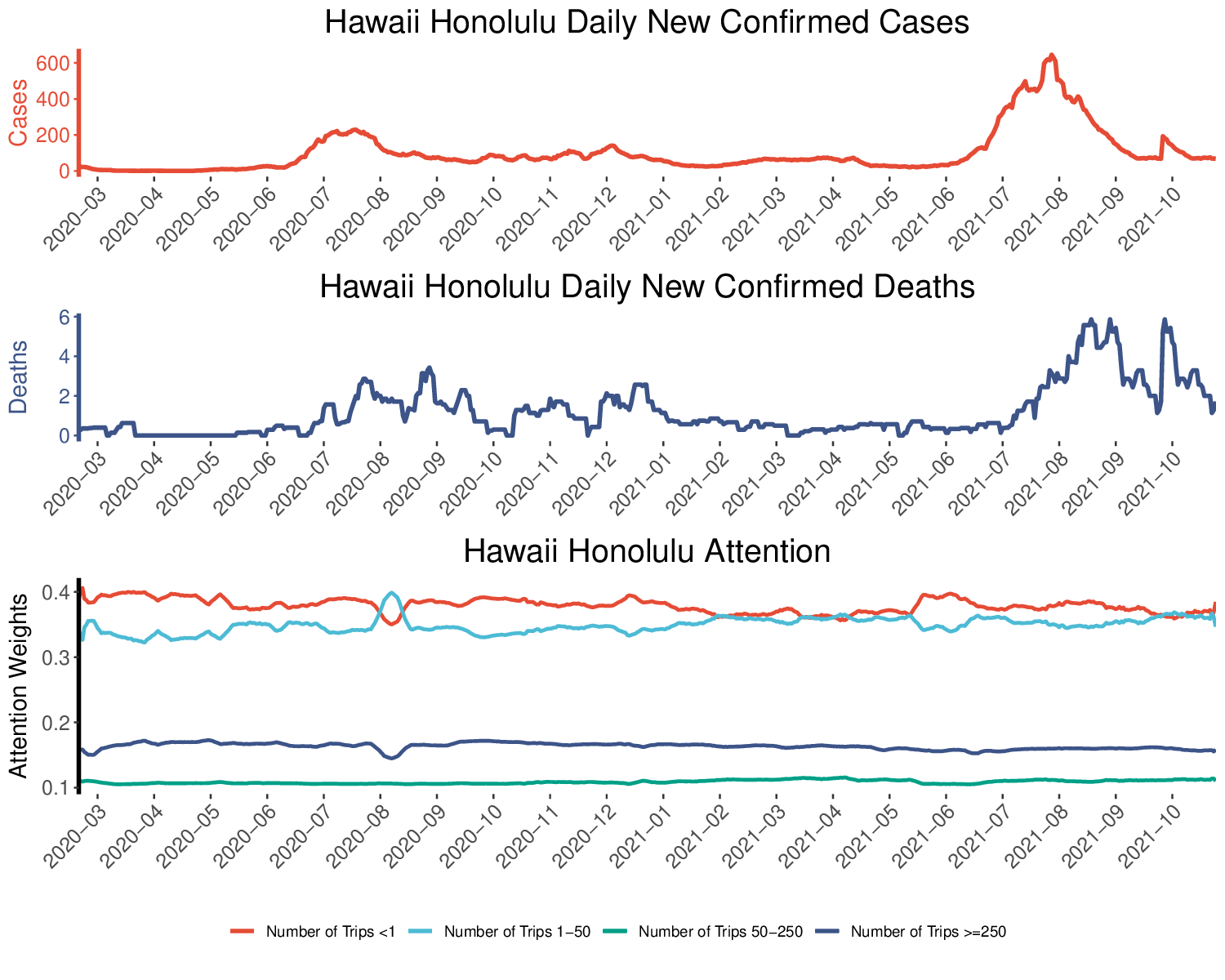}}
	\hfill
	\subfloat[Illinois, Cook County]{\includegraphics[width = 0.45\textwidth]{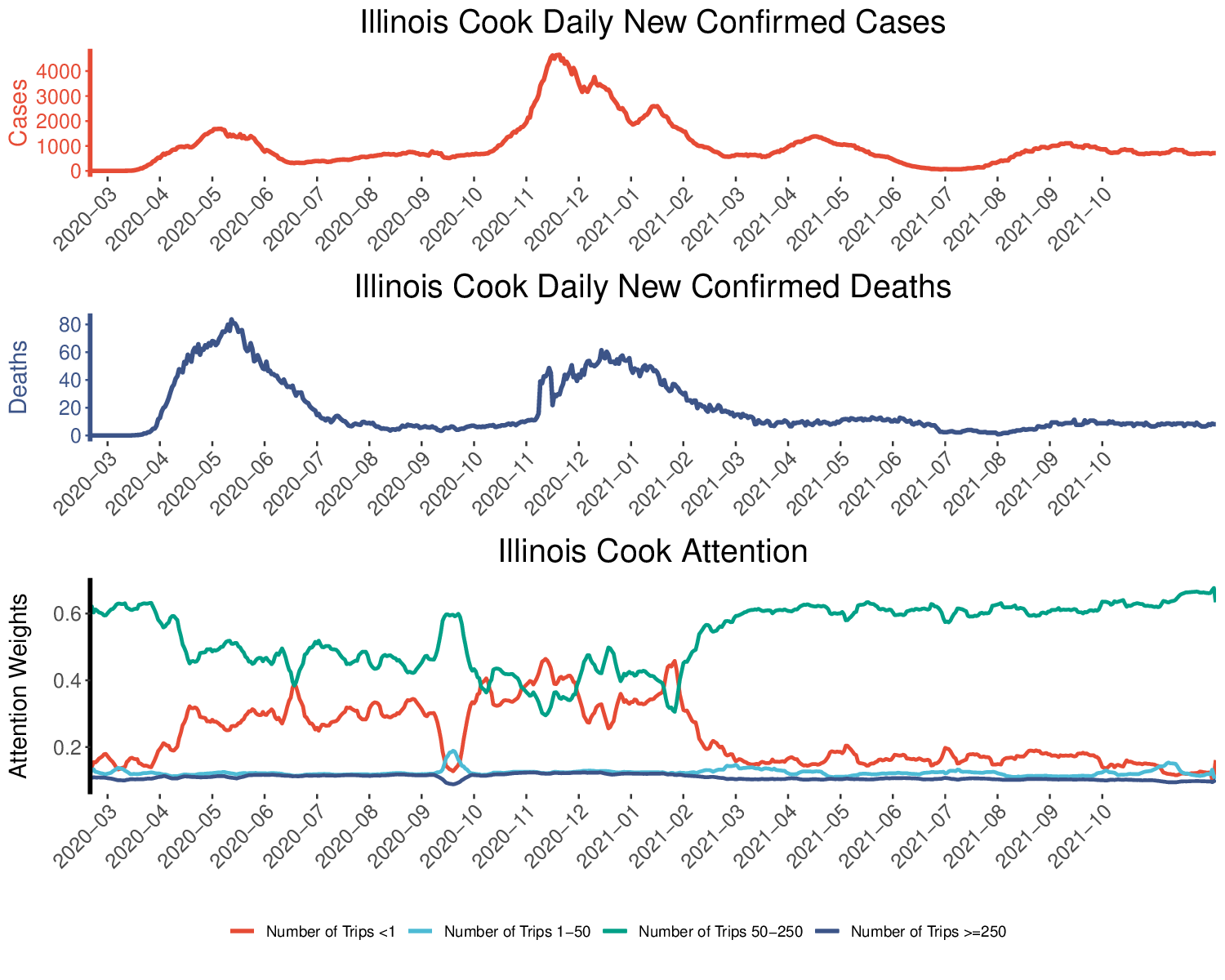}} 
	\hfill
	\subfloat[Florida, Indian River County]{\includegraphics[width = 0.45\textwidth]{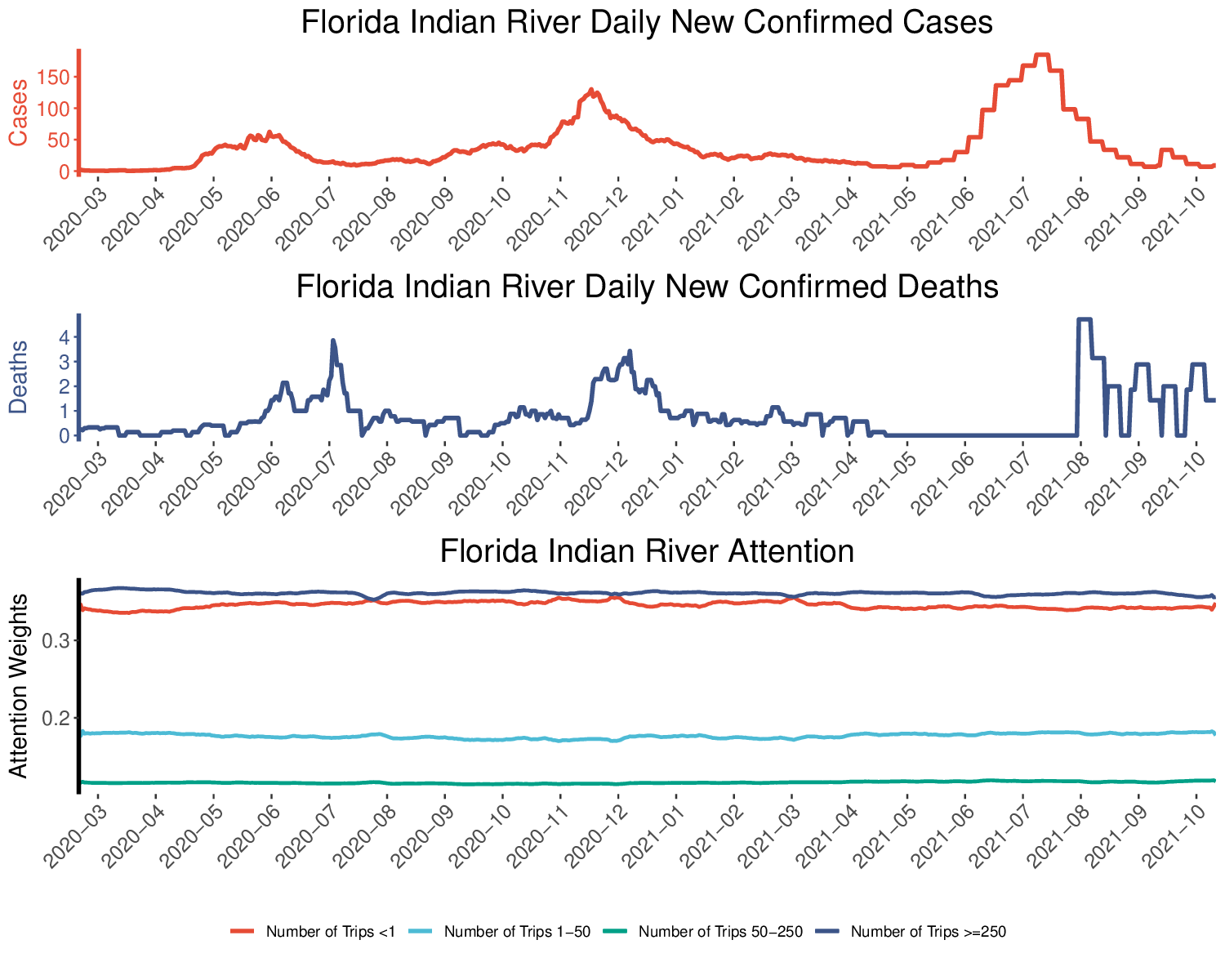}}
\caption{The trajectories of COVID-19 (new cases and new deaths) and the trend of estimated weights for a full range of travel distances across the entire period in California Los Angeles, Hawaii Honolulu, Illinois Cook, and Florida Indian River.}
\label{temporal}
\end{figure*}

Besides the spatial differences, we also seek the temporal distribution of the travel distances. Figure \ref{temporal} illustrates the trajectories of COVID-19 (new cases and new deaths) and the estimated daily weights for a full range of the population taking trips from the first day when the confirmed cases were reported to October 2021 in California Los Angeles, Hawaii Honolulu, Illinois Cook, and Florida Indian River. Honolulu and Indian River counties have quite flatted daily changes of the estimated weights for each class of travel distance. The daily weights for the population taking ``community-level" trips always significantly exceed those of other classes of travel distances and fluctuate at the beginning of the epidemic and after June 2021 in Los Angeles County. In particular, the trajectories of the daily weights for the population taking ``community-level" and ``in-state-level" trips are qualitatively dynamic in Cook County. The values of former ones were over 0.6 at the beginning of the epidemic and decreased to approximately 0.4 after April 2020, and returned over 0.6 after March 2021. In contrast, the values of the later ones began with 0.2 and gradually increased to 0.4, reached their peak around November 2020, and finally decreased to below 0.2 after March 2021.

\subsection{Prediction Performance}
We assess the prediction performance of the S2SEA-Net and compare it with a sequence-to-sequence model using the test data, which is represented by the green bars in Figure \ref{structure}. We calculate the Root Mean Square Error (RMSE) as a metric for comparison between the two models.

\begin{itemize}
\item \textbf{Sequence-to-sequence Model}: The encoder input data is defined as $\operatorname{Concat}(\textbf{X}_{t.k}^{(E)}, D_{t.k})$.
\item \textbf{S2SEA-Net Model}: The encoder input data is defined as $\operatorname{Concat}(\textbf{X}_{t.k}^{(E)}, \operatorname{Attention}(D_{t.k}))$.
\end{itemize}

\begin{table}[!htb]
	\centering
	\caption{The average values of $\operatorname{RMSE}_7$ and $\operatorname{RMSE}_{30}$ for two models of all infected counties.}
	\begin{tabular}{ccccc}
		\toprule 
		Models & \multicolumn{2}{c}{$\operatorname{RMSE}_7$} & \multicolumn{2}{c}{$\operatorname{RMSE}_{30}$} \\
		\cmidrule{2-5}
		& Cases & Deaths & Cases & Deaths \\
		\midrule 
		Sequence-to-sequence & 433.423 &  0.416 & 1281.237 & 0.631 \\
		S2SEA-Net & 354.029 & 0.412 & 852.085 & 0.529 \\
	    \bottomrule
	\end{tabular}
    \label{table1}
\end{table}
Table \ref{table1} provides a summary of the average Root Mean Square Errors (RMSEs) for all 3118 infected counties. Both $\operatorname{RMSE}_7$ and $\operatorname{RMSE}_{30}$ for new cases and new deaths demonstrate superior performance of the S2SEA-Net over the sequence-to-sequence only model.       

\begin{figure*}[!htb]
\centering
	\subfloat[California, Los Angeles County]{\includegraphics[width = 0.45\textwidth]{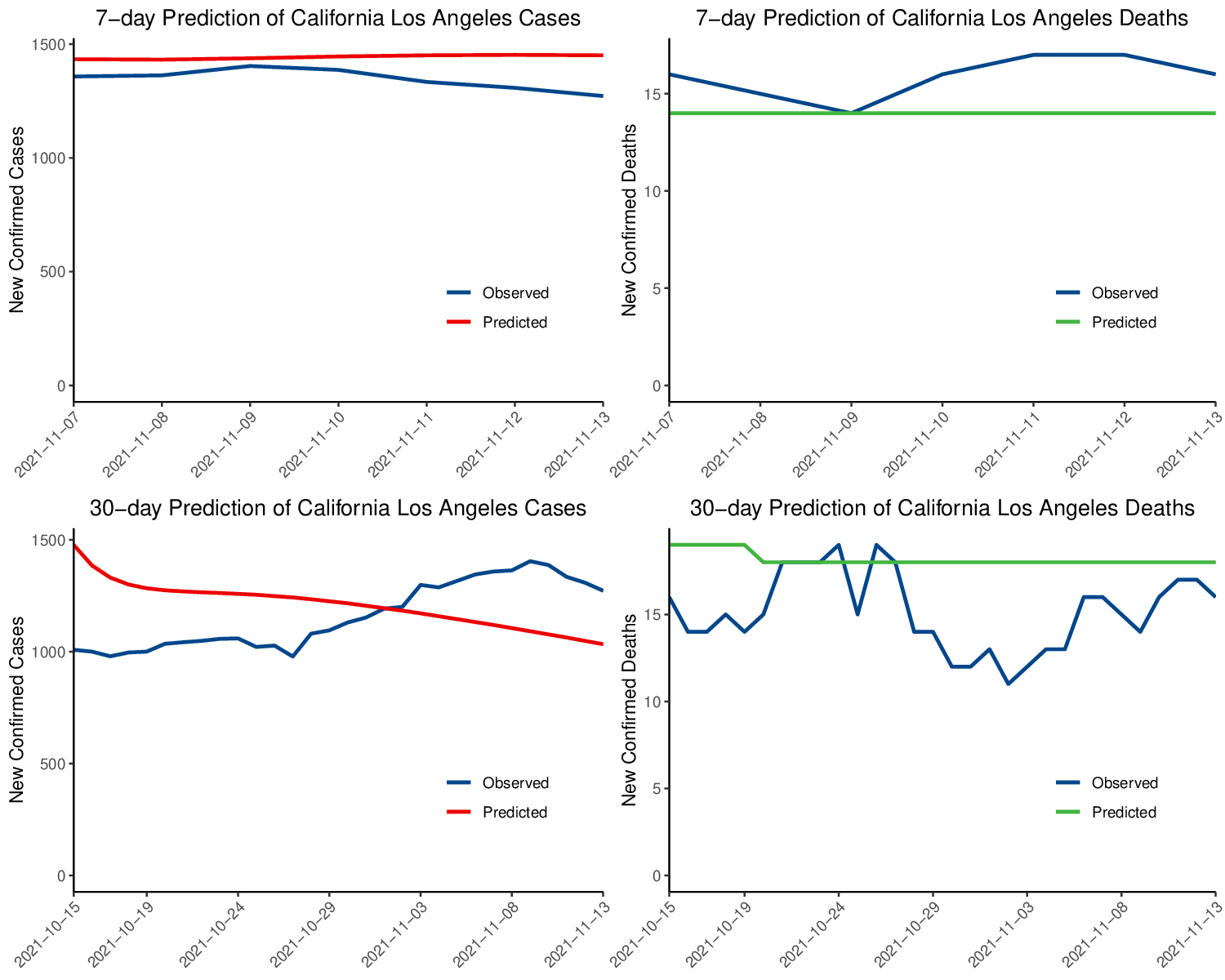}}
	\hfill
	\subfloat[Hawaii, Honolulu County]{\includegraphics[width = 0.45\textwidth]{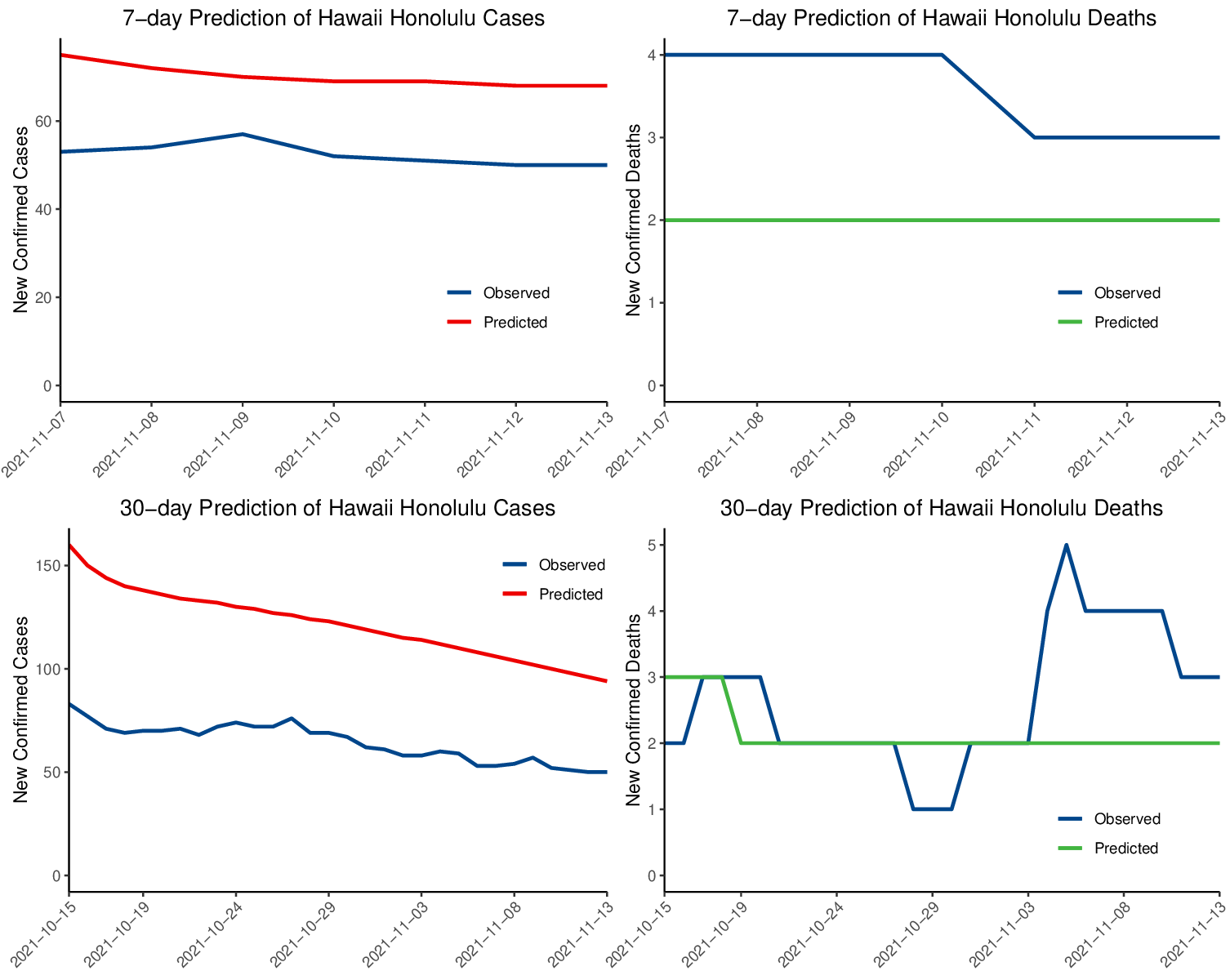}}
	\hfill
	\subfloat[Illinois, Cook County]{\includegraphics[width = 0.45\textwidth]{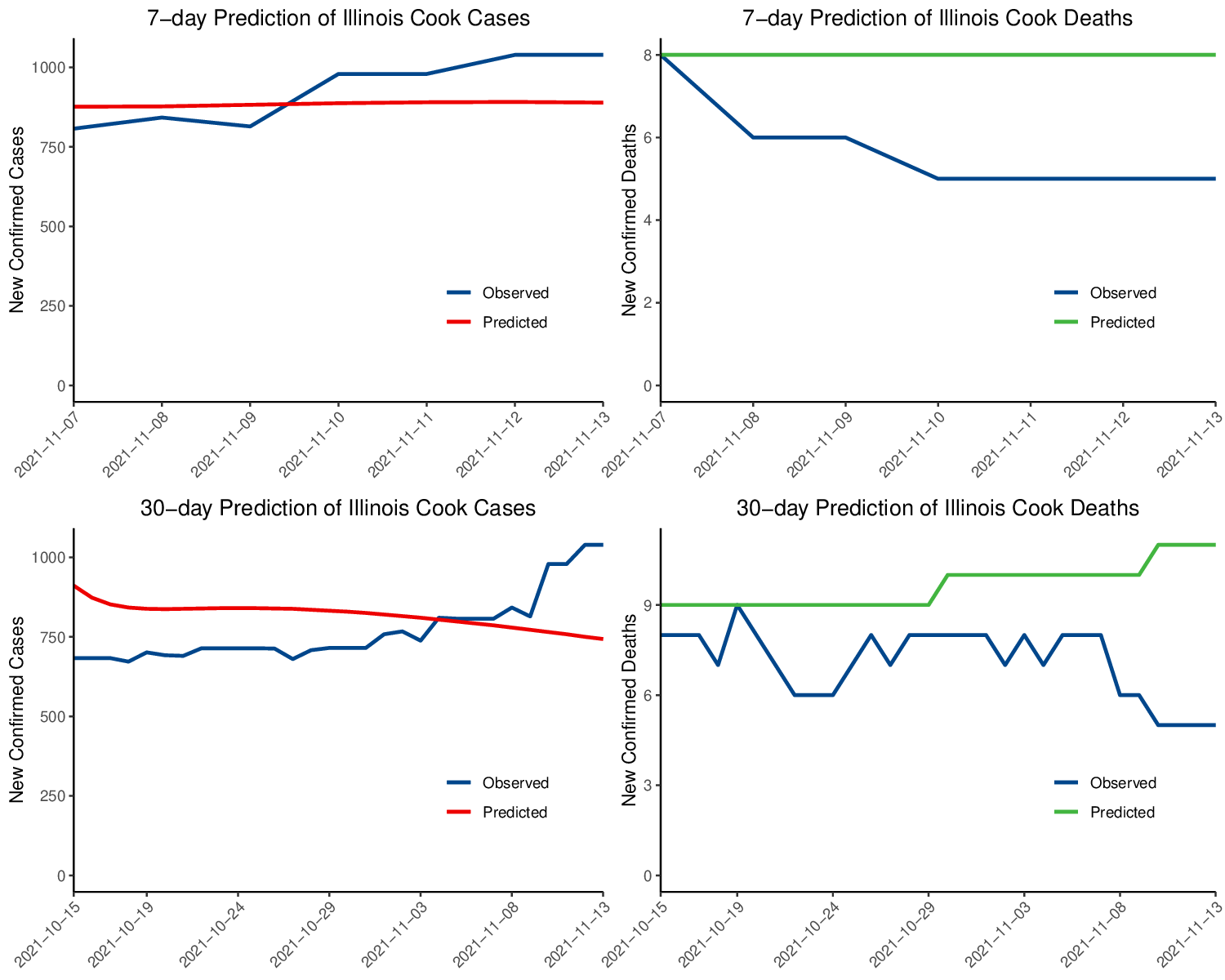}} 
	\hfill
	\subfloat[Florida, Indian River County]{\includegraphics[width = 0.45\textwidth]{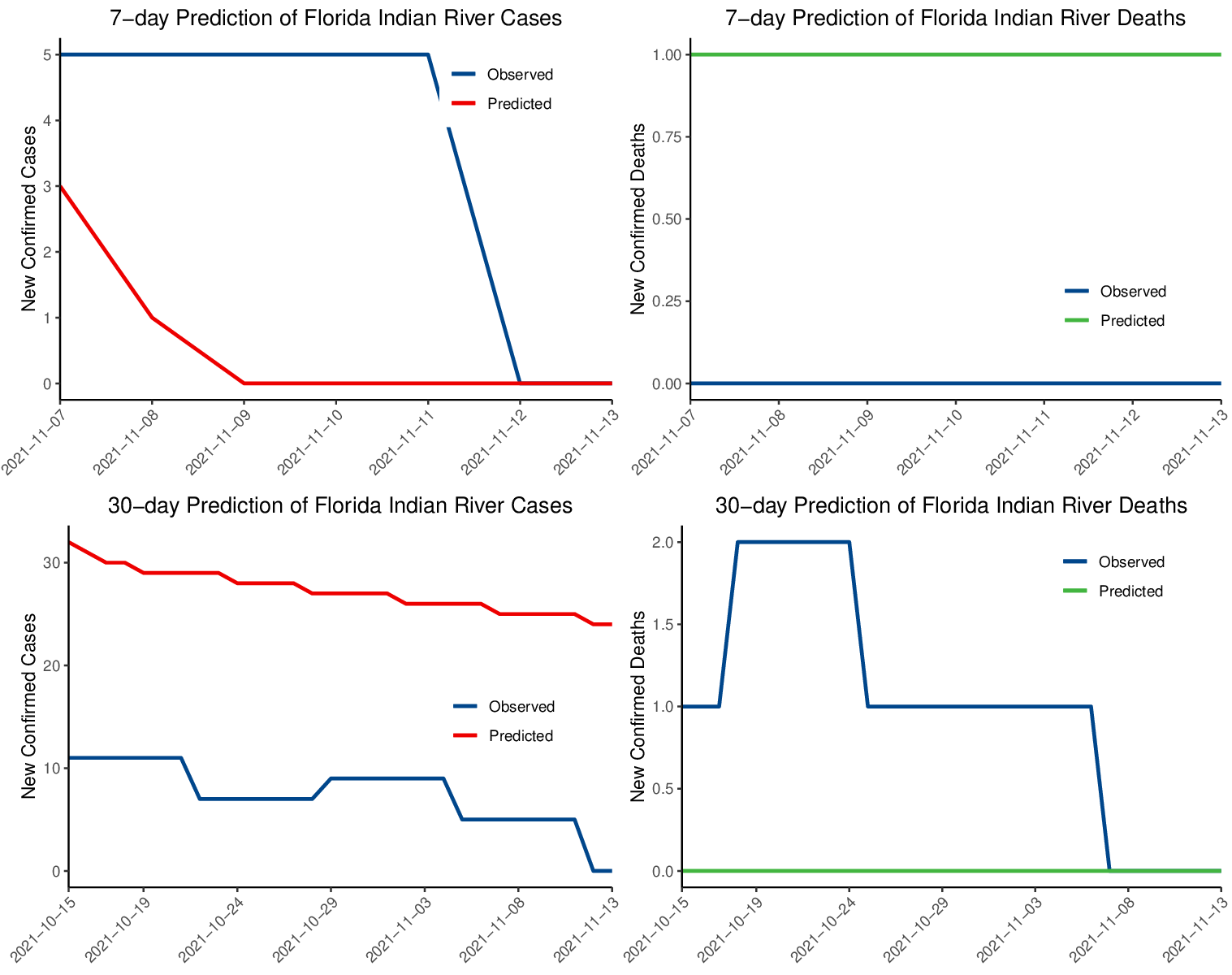}} 
\caption{7- and 30- day new confirmed cases and new deaths prediction of California Los Angeles, Hawaii Honolulu, Illinois Cook, and Florida Indian River.}
\label{predict}
\end{figure*}

To further illustrate the county-level prediction performance of the S2SEA-Net, we display the last 7- and 30-day predicted and observed new cases and new deaths before November 13, 2021, for four representative counties in Figure \ref{predict}. These counties were selected based on their relatively larger average weights over the entire period for the population taking trips for each of the four classes of travel distances (See Table \ref{table2}). Additionally, these counties exhibit relatively high population densities. Figure \ref{predict} clearly illustrates that the gaps between predicted and observed new cases and new deaths are very small for California Los Angeles, Hawaii Honolulu, Illinois Cook, and Florida Indian River.

\section{Discussion}
Our S2SEA-Net model contributes to two significant aspects of COVID-19 analysis. 
Not only does it effectively predict the county-level sequential 7- and 30-day new cases and deaths, but it also estimates the time-varying weights associated with different classes of travel distances, namely ``community-level", ``county-level", ``in-state-level", and ``out-state-level" trips. We have incorporated these features into a deep learning model with an added attention module to improve the accuracy of predicting new confirmed cases and deaths across infected counties in the United States. By incorporating travel data, our model provides valuable insights into the factors influencing the spread of the epidemic.

Our analysis spans various classes of travel distances, from the early stages of the COVID-19 pandemic to November 2021. We observed a decrease in the number of people taking trips in the initial stages of the pandemic, likely due to the implementation of COVID-19 interventions. However, as restrictions eased and businesses reopened, there was a gradual increase in short and intermediate travel distances. This suggests that people began resuming their routine trips, particularly ``community-level" travel distances. This is reflected in the time-varying attention weights, with ``community-level" trips dominating in Los Angeles County, California, after May 2020.

Additionally, regional travel behavior changes emerged as a result of the pandemic. Some individuals opted for shorter-distance travel during the summer of 2020, a trend observed in Illinois and supported by a survey indicating that people were willing to travel shorter distances for vacations. This voluntary shift in travel behavior is reflected in the increase in ``in-state-level" trips and higher attention weights in Cook County, Illinois.

We also found spatial variations in travel distance patterns and their impact on COVID-19 spread. These differences may be attributed to shifts in travel preferences, such as an increase in car travel and a decrease in long-distance travel options like flights. The weights associated with ``out-state-level" trips are generally lower than those for shorter-distance trips. However, certain regions, such as Florida, exhibit relatively higher weights for ``out-state-level" trips, likely due to the tourism industry's influence.

Temporal patterns in travel distances revealed flattened weight curves in regions with less severe COVID-19 spread. Notably, in Cook County, Illinois, when COVID-19 cases surged, attention shifted towards ``community-level" trips over ``in-state-level" trips in late 2020.

S2SEA-Net effectively captures the influence of people's travel distances on the spatio-temporal dynamics of the epidemic. In fact, S2SEA-Net goes beyond analyzing the impact of travel distances on epidemic indicators. In addition, it possesses the capability to explore correlations among spatio-temporal sequences. The model's estimated weights are highly interpretable, facilitating the opening of the black box of deep learning. The versatility of S2SEA-Net allows for its broader application, given its proficiency in examining the interplay between spatio-temporal trends, thereby enhancing its utility in various contexts.

\section{Conclusion}
Understanding the relationship between travel patterns and the spread of COVID-19 is crucial for effective pandemic control. By identifying the weights associated with different classes of travel distances, our model offers insights that can aid policymakers in mitigating the ongoing spread of COVID-19. Specifically, it suggests that targeted interventions to reduce travel within specific distance categories may help curb the epidemic. Moreover, S2SEA-Net, characterized by its high precision and spatio-temporal interpretability, effectively reveals the interrelationships between sequences and possesses wide-ranging applications across various domains in complex data analysis.

\section*{Data availability}
The data used in this study are publicly available from reputable sources: COVID-19 data can be accessed from The New York Times at the following link: \url{https://github.com/nytimes/covid-19-data}. Travel data for different classes of travel distances is available from the Bureau of Transportation Statistics at: \url{https://www.bts.gov/daily-travel}.

\clearpage
\newpage

%\begin{appendices}
%
%\section{Section title of first appendix}\label{secA1}
%
%An appendix contains supplementary information that is not an essential part of the text itself but which may be helpful in providing a more comprehensive understanding of the research problem or it is information that is too cumbersome to be included in the body of the paper.
%
%%%=============================================%%
%%% For submissions to Nature Portfolio Journals %%
%%% please use the heading ``Extended Data''.   %%
%%%=============================================%%
%
%%%=============================================================%%
%%% Sample for another appendix section			       %%
%%%=============================================================%%
%
%%% \section{Example of another appendix section}\label{secA2}%
%%% Appendices may be used for helpful, supporting or essential material that would otherwise 
%%% clutter, break up or be distracting to the text. Appendices can consist of sections, figures, 
%%% tables and equations etc.
%
%\end{appendices}


\bibliography{sn-bibliography}% common bib file


\begin{thebibliography}{27}
\ifx \bisbn   \undefined \def \bisbn  #1{ISBN #1}\fi
\ifx \binits  \undefined \def \binits#1{#1}\fi
\ifx \bauthor  \undefined \def \bauthor#1{#1}\fi
\ifx \batitle  \undefined \def \batitle#1{#1}\fi
\ifx \bjtitle  \undefined \def \bjtitle#1{#1}\fi
\ifx \bvolume  \undefined \def \bvolume#1{\textbf{#1}}\fi
\ifx \byear  \undefined \def \byear#1{#1}\fi
\ifx \bissue  \undefined \def \bissue#1{#1}\fi
\ifx \bfpage  \undefined \def \bfpage#1{#1}\fi
\ifx \blpage  \undefined \def \blpage #1{#1}\fi
\ifx \burl  \undefined \def \burl#1{\textsf{#1}}\fi
\ifx \doiurl  \undefined \def \doiurl#1{\url{https://doi.org/#1}}\fi
\ifx \betal  \undefined \def \betal{\textit{et al.}}\fi
\ifx \binstitute  \undefined \def \binstitute#1{#1}\fi
\ifx \binstitutionaled  \undefined \def \binstitutionaled#1{#1}\fi
\ifx \bctitle  \undefined \def \bctitle#1{#1}\fi
\ifx \beditor  \undefined \def \beditor#1{#1}\fi
\ifx \bpublisher  \undefined \def \bpublisher#1{#1}\fi
\ifx \bbtitle  \undefined \def \bbtitle#1{#1}\fi
\ifx \bedition  \undefined \def \bedition#1{#1}\fi
\ifx \bseriesno  \undefined \def \bseriesno#1{#1}\fi
\ifx \blocation  \undefined \def \blocation#1{#1}\fi
\ifx \bsertitle  \undefined \def \bsertitle#1{#1}\fi
\ifx \bsnm \undefined \def \bsnm#1{#1}\fi
\ifx \bsuffix \undefined \def \bsuffix#1{#1}\fi
\ifx \bparticle \undefined \def \bparticle#1{#1}\fi
\ifx \barticle \undefined \def \barticle#1{#1}\fi
\bibcommenthead
\ifx \bconfdate \undefined \def \bconfdate #1{#1}\fi
\ifx \botherref \undefined \def \botherref #1{#1}\fi
\ifx \url \undefined \def \url#1{\textsf{#1}}\fi
\ifx \bchapter \undefined \def \bchapter#1{#1}\fi
\ifx \bbook \undefined \def \bbook#1{#1}\fi
\ifx \bcomment \undefined \def \bcomment#1{#1}\fi
\ifx \oauthor \undefined \def \oauthor#1{#1}\fi
\ifx \citeauthoryear \undefined \def \citeauthoryear#1{#1}\fi
\ifx \endbibitem  \undefined \def \endbibitem {}\fi
\ifx \bconflocation  \undefined \def \bconflocation#1{#1}\fi
\ifx \arxivurl  \undefined \def \arxivurl#1{\textsf{#1}}\fi
\csname PreBibitemsHook\endcsname

%%% 1
\bibitem[\protect\citeauthoryear{Frankish}{2003}]{frankish2003death}
\begin{barticle}
\bauthor{\bsnm{Frankish}, \binits{H.}}:
\batitle{Death toll continues to climb in congo ebola outbreak}.
\bjtitle{The Lancet}
\bvolume{361}(\bissue{9362}),
\bfpage{1020}
(\byear{2003})
\end{barticle}
\endbibitem

%%% 2
\bibitem[\protect\citeauthoryear{Daszak et~al.}{2000}]{daszak2000emerging}
\begin{barticle}
\bauthor{\bsnm{Daszak}, \binits{P.}},
\bauthor{\bsnm{Cunningham}, \binits{A.A.}},
\bauthor{\bsnm{Hyatt}, \binits{A.D.}}:
\batitle{Emerging infectious diseases of wildlife--threats to biodiversity and
  human health}.
\bjtitle{science}
\bvolume{287}(\bissue{5452}),
\bfpage{443}--\blpage{449}
(\byear{2000})
\end{barticle}
\endbibitem

%%% 3
\bibitem[\protect\citeauthoryear{Donnelly
  et~al.}{2003}]{donnelly2003epidemiological}
\begin{barticle}
\bauthor{\bsnm{Donnelly}, \binits{C.A.}},
\bauthor{\bsnm{Ghani}, \binits{A.C.}},
\bauthor{\bsnm{Leung}, \binits{G.M.}},
\bauthor{\bsnm{Hedley}, \binits{A.J.}},
\bauthor{\bsnm{Fraser}, \binits{C.}},
\bauthor{\bsnm{Riley}, \binits{S.}},
\bauthor{\bsnm{Abu-Raddad}, \binits{L.J.}},
\bauthor{\bsnm{Ho}, \binits{L.-M.}},
\bauthor{\bsnm{Thach}, \binits{T.-Q.}},
\bauthor{\bsnm{Chau}, \binits{P.}}, \betal:
\batitle{Epidemiological determinants of spread of causal agent of severe acute
  respiratory syndrome in hong kong}.
\bjtitle{The lancet}
\bvolume{361}(\bissue{9371}),
\bfpage{1761}--\blpage{1766}
(\byear{2003})
\end{barticle}
\endbibitem

%%% 4
\bibitem[\protect\citeauthoryear{Hall et~al.}{2020}]{hall2020pandemics}
\begin{barticle}
\bauthor{\bsnm{Hall}, \binits{C.M.}},
\bauthor{\bsnm{Scott}, \binits{D.}},
\bauthor{\bsnm{G{\"o}ssling}, \binits{S.}}:
\batitle{Pandemics, transformations and tourism: Be careful what you wish for}.
\bjtitle{Tourism geographies}
\bvolume{22}(\bissue{3}),
\bfpage{577}--\blpage{598}
(\byear{2020})
\end{barticle}
\endbibitem

%%% 5
\bibitem[\protect\citeauthoryear{{WHO~Director~General}}{2020}]{WHO}
\begin{botherref}
\oauthor{\bsnm{{WHO~Director~General}}}:
WHO Director-General's opening remarks at the media briefing on COVID-19 -11
  March 2020.
\url{https://www.who.int/director-general/speeches/detail/who-director-general-s-opening-remarks-at-the-media-briefing-on-covid-19---11-march-2020}
(2020)
\end{botherref}
\endbibitem

%%% 6
\bibitem[\protect\citeauthoryear{{The~New~York~Times}}{2021}]{NYT}
\begin{botherref}
\oauthor{\bsnm{{The~New~York~Times}}}:
Coronavirus in the U.S.: Latest Map and Case Count.
\url{https://www.nytimes.com/interactive/2021/us/covid-cases.html}
(2021)
\end{botherref}
\endbibitem

%%% 7
\bibitem[\protect\citeauthoryear{Tian
  et~al.}{2021}]{tian_tan_jiang_wang_zhang_2021}
\begin{barticle}
\bauthor{\bsnm{Tian}, \binits{T.}},
\bauthor{\bsnm{Tan}, \binits{J.}},
\bauthor{\bsnm{Jiang}, \binits{Y.}},
\bauthor{\bsnm{Wang}, \binits{X.}},
\bauthor{\bsnm{Zhang}, \binits{H.}}:
\batitle{Evaluate the risk of resumption of business for the states of new
  york, new jersey and connecticut via a pre-symptomatic and asymptomatic
  transmission model of covid-19}.
\bjtitle{Journal of Data Science}
\bvolume{19}(\bissue{2}),
\bfpage{178}--\blpage{196}
(\byear{2021})
\doiurl{10.6339/21-JDS994}
\end{barticle}
\endbibitem

%%% 8
\bibitem[\protect\citeauthoryear{Chen and Steiner}{2022}]{chen2022longitudinal}
\begin{barticle}
\bauthor{\bsnm{Chen}, \binits{K.}},
\bauthor{\bsnm{Steiner}, \binits{R.}}:
\batitle{Longitudinal and spatial analysis of americans’ travel distances
  following covid-19}.
\bjtitle{Transportation Research Part D: Transport and Environment}
\bvolume{110},
\bfpage{103414}
(\byear{2022})
\end{barticle}
\endbibitem

%%% 9
\bibitem[\protect\citeauthoryear{Flaxman et~al.}{2020}]{flaxman2020estimating}
\begin{barticle}
\bauthor{\bsnm{Flaxman}, \binits{S.}},
\bauthor{\bsnm{Mishra}, \binits{S.}},
\bauthor{\bsnm{Gandy}, \binits{A.}},
\bauthor{\bsnm{Unwin}, \binits{H.J.T.}},
\bauthor{\bsnm{Mellan}, \binits{T.A.}},
\bauthor{\bsnm{Coupland}, \binits{H.}},
\bauthor{\bsnm{Whittaker}, \binits{C.}},
\bauthor{\bsnm{Zhu}, \binits{H.}},
\bauthor{\bsnm{Berah}, \binits{T.}},
\bauthor{\bsnm{Eaton}, \binits{J.W.}}, \betal:
\batitle{Estimating the effects of non-pharmaceutical interventions on covid-19
  in europe}.
\bjtitle{Nature}
\bvolume{584}(\bissue{7820}),
\bfpage{257}--\blpage{261}
(\byear{2020})
\end{barticle}
\endbibitem

%%% 10
\bibitem[\protect\citeauthoryear{Dehning et~al.}{2020}]{dehning2020inferring}
\begin{botherref}
\oauthor{\bsnm{Dehning}, \binits{J.}},
\oauthor{\bsnm{Zierenberg}, \binits{J.}},
\oauthor{\bsnm{Spitzner}, \binits{F.P.}},
\oauthor{\bsnm{Wibral}, \binits{M.}},
\oauthor{\bsnm{Neto}, \binits{J.P.}},
\oauthor{\bsnm{Wilczek}, \binits{M.}},
\oauthor{\bsnm{Priesemann}, \binits{V.}}:
Inferring change points in the spread of covid-19 reveals the effectiveness of
  interventions.
Science
\textbf{369}(6500)
(2020)
\end{botherref}
\endbibitem

%%% 11
\bibitem[\protect\citeauthoryear{Jia et~al.}{2020}]{jia2020population}
\begin{barticle}
\bauthor{\bsnm{Jia}, \binits{J.S.}},
\bauthor{\bsnm{Lu}, \binits{X.}},
\bauthor{\bsnm{Yuan}, \binits{Y.}},
\bauthor{\bsnm{Xu}, \binits{G.}},
\bauthor{\bsnm{Jia}, \binits{J.}},
\bauthor{\bsnm{Christakis}, \binits{N.A.}}:
\batitle{Population flow drives spatio-temporal distribution of covid-19 in
  china}.
\bjtitle{Nature}
\bvolume{582}(\bissue{7812}),
\bfpage{389}--\blpage{394}
(\byear{2020})
\end{barticle}
\endbibitem

%%% 12
\bibitem[\protect\citeauthoryear{Chinazzi et~al.}{2020}]{chinazzi2020effect}
\begin{barticle}
\bauthor{\bsnm{Chinazzi}, \binits{M.}},
\bauthor{\bsnm{Davis}, \binits{J.T.}},
\bauthor{\bsnm{Ajelli}, \binits{M.}},
\bauthor{\bsnm{Gioannini}, \binits{C.}},
\bauthor{\bsnm{Litvinova}, \binits{M.}},
\bauthor{\bsnm{Merler}, \binits{S.}},
\bauthor{\bsnm{Piontti}, \binits{A.P.}},
\bauthor{\bsnm{Mu}, \binits{K.}},
\bauthor{\bsnm{Rossi}, \binits{L.}},
\bauthor{\bsnm{Sun}, \binits{K.}}, \betal:
\batitle{The effect of travel restrictions on the spread of the 2019 novel
  coronavirus (covid-19) outbreak}.
\bjtitle{Science}
\bvolume{368}(\bissue{6489}),
\bfpage{395}--\blpage{400}
(\byear{2020})
\end{barticle}
\endbibitem

%%% 13
\bibitem[\protect\citeauthoryear{Schlosser et~al.}{2020}]{schlosser2020covid}
\begin{barticle}
\bauthor{\bsnm{Schlosser}, \binits{F.}},
\bauthor{\bsnm{Maier}, \binits{B.F.}},
\bauthor{\bsnm{Jack}, \binits{O.}},
\bauthor{\bsnm{Hinrichs}, \binits{D.}},
\bauthor{\bsnm{Zachariae}, \binits{A.}},
\bauthor{\bsnm{Brockmann}, \binits{D.}}:
\batitle{Covid-19 lockdown induces disease-mitigating structural changes in
  mobility networks}.
\bjtitle{Proceedings of the National Academy of Sciences}
\bvolume{117}(\bissue{52}),
\bfpage{32883}--\blpage{32890}
(\byear{2020})
\end{barticle}
\endbibitem

%%% 14
\bibitem[\protect\citeauthoryear{Galeazzi et~al.}{2021}]{galeazzi2021human}
\begin{barticle}
\bauthor{\bsnm{Galeazzi}, \binits{A.}},
\bauthor{\bsnm{Cinelli}, \binits{M.}},
\bauthor{\bsnm{Bonaccorsi}, \binits{G.}},
\bauthor{\bsnm{Pierri}, \binits{F.}},
\bauthor{\bsnm{Schmidt}, \binits{A.L.}},
\bauthor{\bsnm{Scala}, \binits{A.}},
\bauthor{\bsnm{Pammolli}, \binits{F.}},
\bauthor{\bsnm{Quattrociocchi}, \binits{W.}}:
\batitle{Human mobility in response to covid-19 in france, italy and uk}.
\bjtitle{Scientific Reports}
\bvolume{11}(\bissue{1}),
\bfpage{1}--\blpage{10}
(\byear{2021})
\end{barticle}
\endbibitem

%%% 15
\bibitem[\protect\citeauthoryear{Ronneberger et~al.}{2015}]{ronneberger2015u}
\begin{bchapter}
\bauthor{\bsnm{Ronneberger}, \binits{O.}},
\bauthor{\bsnm{Fischer}, \binits{P.}},
\bauthor{\bsnm{Brox}, \binits{T.}}:
\bctitle{U-net: Convolutional networks for biomedical image segmentation}.
In: \bbtitle{International Conference on Medical Image Computing and
  Computer-assisted Intervention},
pp. \bfpage{234}--\blpage{241}
(\byear{2015}).
\bcomment{Springer}
\end{bchapter}
\endbibitem

%%% 16
\bibitem[\protect\citeauthoryear{Ren et~al.}{2015}]{ren2015faster}
\begin{botherref}
\oauthor{\bsnm{Ren}, \binits{S.}},
\oauthor{\bsnm{He}, \binits{K.}},
\oauthor{\bsnm{Girshick}, \binits{R.}},
\oauthor{\bsnm{Sun}, \binits{J.}}:
Faster r-cnn: Towards real-time object detection with region proposal networks.
arXiv preprint arXiv:1506.01497
(2015)
\end{botherref}
\endbibitem

%%% 17
\bibitem[\protect\citeauthoryear{Hochreiter and
  Schmidhuber}{1997}]{hochreiter1997long}
\begin{barticle}
\bauthor{\bsnm{Hochreiter}, \binits{S.}},
\bauthor{\bsnm{Schmidhuber}, \binits{J.}}:
\batitle{Long short-term memory}.
\bjtitle{Neural computation}
\bvolume{9}(\bissue{8}),
\bfpage{1735}--\blpage{1780}
(\byear{1997})
\end{barticle}
\endbibitem

%%% 18
\bibitem[\protect\citeauthoryear{Sutskever
  et~al.}{2014}]{sutskever2014sequence}
\begin{bchapter}
\bauthor{\bsnm{Sutskever}, \binits{I.}},
\bauthor{\bsnm{Vinyals}, \binits{O.}},
\bauthor{\bsnm{Le}, \binits{Q.V.}}:
\bctitle{Sequence to sequence learning with neural networks}.
In: \bbtitle{Advances in Neural Information Processing Systems},
pp. \bfpage{3104}--\blpage{3112}
(\byear{2014})
\end{bchapter}
\endbibitem

%%% 19
\bibitem[\protect\citeauthoryear{Pang et~al.}{2021}]{pang2021collaborative}
\begin{barticle}
\bauthor{\bsnm{Pang}, \binits{J.}},
\bauthor{\bsnm{Huang}, \binits{Y.}},
\bauthor{\bsnm{Xie}, \binits{Z.}},
\bauthor{\bsnm{Li}, \binits{J.}},
\bauthor{\bsnm{Cai}, \binits{Z.}}:
\batitle{Collaborative city digital twin for the covid-19 pandemic: A federated
  learning solution}.
\bjtitle{Tsinghua science and technology}
\bvolume{26}(\bissue{5}),
\bfpage{759}--\blpage{771}
(\byear{2021})
\end{barticle}
\endbibitem

%%% 20
\bibitem[\protect\citeauthoryear{Zhang-James et~al.}{2021}]{zhang2021seq2seq}
\begin{botherref}
\oauthor{\bsnm{Zhang-James}, \binits{Y.}},
\oauthor{\bsnm{Hess}, \binits{J.}},
\oauthor{\bsnm{Salekin}, \binits{A.}},
\oauthor{\bsnm{Wang}, \binits{D.}},
\oauthor{\bsnm{Chen}, \binits{S.}},
\oauthor{\bsnm{Winkelstein}, \binits{P.}},
\oauthor{\bsnm{Morley}, \binits{C.P.}},
\oauthor{\bsnm{Faraone}, \binits{S.V.}}:
A seq2seq model to forecast the covid-19 cases, deaths and reproductive r
  numbers in us counties
(2021)
\end{botherref}
\endbibitem

%%% 21
\bibitem[\protect\citeauthoryear{Tian et~al.}{2021}]{Tian2020}
\begin{barticle}
\bauthor{\bsnm{Tian}, \binits{T.}},
\bauthor{\bsnm{Jiang}, \binits{Y.}},
\bauthor{\bsnm{Zhang}, \binits{Y.}},
\bauthor{\bsnm{Li}, \binits{Z.}},
\bauthor{\bsnm{Wang}, \binits{X.}},
\bauthor{\bsnm{Zhang}, \binits{H.}}:
\batitle{Covinet: A deep learning-based and interpretable prediction model for
  the county-wise trajectories of covid-19 in the united states}.
\bjtitle{medRxiv}
(\byear{2021})
\doiurl{10.1101/2020.05.26.20113787}
{\href{https://arxiv.org/abs/https://www.medrxiv.org/content/early/2020/05/27/2020.05.26.20113787.full.pdf}{{https://www.medrxiv.org/content/early/2020/05/27/2020.05.26.20113787.full.pdf}}}
\end{barticle}
\endbibitem

%%% 22
\bibitem[\protect\citeauthoryear{Alassafi et~al.}{2022}]{alassafi2022time}
\begin{barticle}
\bauthor{\bsnm{Alassafi}, \binits{M.O.}},
\bauthor{\bsnm{Jarrah}, \binits{M.}},
\bauthor{\bsnm{Alotaibi}, \binits{R.}}:
\batitle{Time series predicting of covid-19 based on deep learning}.
\bjtitle{Neurocomputing}
\bvolume{468},
\bfpage{335}--\blpage{344}
(\byear{2022})
\end{barticle}
\endbibitem

%%% 23
\bibitem[\protect\citeauthoryear{Chandra et~al.}{2022}]{chandra2022deep}
\begin{barticle}
\bauthor{\bsnm{Chandra}, \binits{R.}},
\bauthor{\bsnm{Jain}, \binits{A.}},
\bauthor{\bsnm{Singh~Chauhan}, \binits{D.}}:
\batitle{Deep learning via lstm models for covid-19 infection forecasting in
  india}.
\bjtitle{PloS one}
\bvolume{17}(\bissue{1}),
\bfpage{0262708}
(\byear{2022})
\end{barticle}
\endbibitem

%%% 24
\bibitem[\protect\citeauthoryear{Sinha et~al.}{2022}]{sinha2022analysis}
\begin{bchapter}
\bauthor{\bsnm{Sinha}, \binits{T.}},
\bauthor{\bsnm{Chowdhury}, \binits{T.}},
\bauthor{\bsnm{Shaw}, \binits{R.N.}},
\bauthor{\bsnm{Ghosh}, \binits{A.}}:
\bctitle{Analysis and prediction of covid-19 confirmed cases using deep
  learning models: A comparative study}.
In: \bbtitle{Advanced Computing and Intelligent Technologies},
pp. \bfpage{207}--\blpage{218}.
\bpublisher{Springer}, \blocation{???}
(\byear{2022})
\end{bchapter}
\endbibitem

%%% 25
\bibitem[\protect\citeauthoryear{Giuliani et~al.}{2020}]{giuliani2020modelling}
\begin{barticle}
\bauthor{\bsnm{Giuliani}, \binits{D.}},
\bauthor{\bsnm{Dickson}, \binits{M.M.}},
\bauthor{\bsnm{Espa}, \binits{G.}},
\bauthor{\bsnm{Santi}, \binits{F.}}:
\batitle{Modelling and predicting the spatio-temporal spread of covid-19 in
  italy}.
\bjtitle{BMC infectious diseases}
\bvolume{20}(\bissue{1}),
\bfpage{1}--\blpage{10}
(\byear{2020})
\end{barticle}
\endbibitem

%%% 26
\bibitem[\protect\citeauthoryear{Fukui et~al.}{2019}]{fukui2019attention}
\begin{bchapter}
\bauthor{\bsnm{Fukui}, \binits{H.}},
\bauthor{\bsnm{Hirakawa}, \binits{T.}},
\bauthor{\bsnm{Yamashita}, \binits{T.}},
\bauthor{\bsnm{Fujiyoshi}, \binits{H.}}:
\bctitle{Attention branch network: Learning of attention mechanism for visual
  explanation}.
In: \bbtitle{Proceedings of the IEEE/CVF Conference on Computer Vision and
  Pattern Recognition},
pp. \bfpage{10705}--\blpage{10714}
(\byear{2019})
\end{bchapter}
\endbibitem

%%% 27
\bibitem[\protect\citeauthoryear{{National~Oceanic~And~Atmospheric~Administration}}{2022}]{Counties}
\begin{botherref}
\oauthor{\bsnm{{National~Oceanic~And~Atmospheric~Administration}}}:
U.S. Counties.
\url{https://www.weather.gov/gis/Counties}
(2022)
\end{botherref}
\endbibitem

\end{thebibliography}

\end{document}